\documentclass[epj]{svjour}
\usepackage[dvips]{graphicx}
\usepackage{amsmath,amssymb}
\newcommand{\bce}{\begin{center}}
\newcommand{\ece}{\end{center}}
\newcommand{\be}{\begin{equation}}
\newcommand{\ee}{\end{equation}}
\newcommand{\bea}{\begin{eqnarray}}
\newcommand{\eea}{\end{eqnarray}}
\newcommand{\bdes}{\begin{description}}
\newcommand{\edes}{\end{description}}
\newcommand{\bit}{\begin{itemize}}
\newcommand{\eit}{\end{itemize}}
\newcommand{\btt}{\begin{tt}}
\newcommand{\ett}{\end{tt}}

\def\E{\> = \>}
\def\EA{&=&}
\def\non{\nonumber\\}
\def\To{\> \longrightarrow \>}

\def\vecrho{\mbox{\boldmath$\rho$}}

\def\fb{{\bf b}}  %  f for ``fat'' oder ``fett'' ...
\def\fB{{\bf B}}
\def\fk{{\bf k}}
\def\fK{{\bf K}}
\def\fp{{\bf p}}
\def\fq{{\bf q}}

\def\fv{{\bf v}}

\def\fw{{\bf w}}
\def\fx{{\bf x}}

\def\Iint{\int_{-\infty}^{+\infty} dt}
\def\Def{\> := \>}
\def\deF{\> =: \>}
\def\la{\left\langle \,}
\def\rat{\, \right\rangle_t}
\newcommand{\A}{\mathbb A}
\newcommand{\B}{\mathbb B}
\newcommand{\C}{\mathbb C}
\newcommand{\V}{\mathbb V}

\newcommand{\sgn}{{\rm sgn}}
\renewcommand{\theequation}{\thesection.\arabic{equation}}
\begin{document}
\title{Variational Approximations in a Path-Integral Description of Potential Scattering}

\author{J.~Carron \thanks{\emph{Present address:} Institute for Astronomy, ETH Zurich, 
Wolfgang-Pauli-Strasse 27, CH-8093 Z\"urich} 
\and R.~Rosenfelder\thanks{e-mail: {\tt roland.rosenfelder@psi.ch}}
}
\institute{Particle Theory Group, Paul Scherrer Institute,
CH-5232 Villigen PSI, Switzerland} 
%\and Institute for Astronomy, ETH Zurich, 
%Wolfgang-Pauli-Strasse 27, CH-8093 Z\"urich (present address)} 
%
\date{Published: Eur. Phys. J. A {\bf 45} 193- 215 (2010)}
\abstract{
Using a  recent path integral representation for the ${\cal T}$-matrix in nonrelativistic
potential scattering we investigate new variational approximations in this framework. By means of the
Feynman-Jensen variational principle and the most general ansatz quadratic in the velocity variables
-- over which one has to integrate functionally -- we obtain variational equations which contain 
classical elements (trajectories) as well as quantum-mechanical ones (wave spreading).
We analyse these equations and solve them numerically by iteration, a procedure best suited 
at high energy.
The first correction to the variational result arising from a cumulant expansion 
is also evaluated. Comparison is made with exact partial-wave results for scattering from  a 
Gaussian potential and better agreement is found at large scattering angles where the standard 
eikonal-type approximations fail. 
} %end of abstract
\maketitle
%\vspace{-2cm}
%\hspace{1cm} 
%\vspace{2cm}

%%%%%%%%%%%%%%%%%%%%%%%%%%%%%%%%%%%%%%%%%%%%%%%%%%%%%%%%%%%%%%%%%%%%%%%
%%%  ********************* INTRODUCTION ******************************
%%%%%%%%%%%%%%%%%%%%%%%%%%%%%%%%%%%%%%%%%%%%%%%%%%%%%%%%%%%%%%%%%%%%%%

\section{Introduction}

Variational approaches to quantum mechanical scattering have a long history and are well covered in 
standard text books (e.g. \cite{Mess,GoWat,Tayl,Newt}). Actually, as formulated in ref. \cite{GeRaSp}, 
it is ''... possible to construct systematically a variational principle for just about any given 
quantity of interest, provided that the entities which enter into the definition of (that quantity) 
are uniquely defined by a set of equations ...''. For quantum physics the observables obtained from 
a solution of Schr\"odinger's equation are of particular interest and therefore variational 
principles have been available since the beginning of quantum mechanics.
Best known is the
Rayleigh-Ritz variational principle for the ground-state energy of bound systems
but the continuous spectrum is also accessible to a variational treatment.
Most prominent among the variational principles for scattering are
Schwinger's \cite{LevSchw} and Kohn's \cite{Kohn} expressions. In particular, Kohn's variational 
principle is widely used in nuclear \cite{ViKiRo,KRVMG,LiRoSp1} and atomic \cite{Nesb,CoArPl} 
physics for an approximate description of few-body scattering near thresholds.

While these approaches benefit from the flexibility which ingenious trial wave functions offer 
it is well known that in many-body systems or in field theory the use of {\it wave functions} 
(or functionals) ceases to be useful. The path-integral method 
where one integrates functionally over the degrees of
freedom weighted by the exponential of the classical {\it action} is much more general although the 
cases where one can actually perform the path integral are rare. Therefore, in general, one has to
resort to approximations, such as perturbation theory or brute-force numerical evaluation of the
functional integral on a (space-time-) lattice. If this is not appropriate or feasible, one may
use a variational principle extended to actions. A prime example is the Feynman-Jensen variational 
principle which has been used to obtain the best semi-analytic ground-state energy of an electron 
in an ionic crystal (the polaron problem \cite{Feyn}). 

Oddly enough, scattering has mostly remained outside the path-integral approach
and it is only at zero energy that bounds for the scattering length have been obtained from 
the path integral in the imaginary-time formulation \cite{Gel,GeSp}.
Expanding on previous attempts \cite{CFJM} a real-time path-integral representation
for the nonrelativistic ${\cal T}$-matrix in potential scattering has recently  
been derived \cite{Rose}. In this formulation the particle travels mainly along a simple 
reference path while quantum fluctuations around this
path are taking into account by functional integration over velocities. It has been shown that
this description gives the exact Born series to all orders if an expansion in powers of the 
potential is done and reduces to the eikonal approximation (valid at high energies and small 
scattering angles) when the quantum fluctuations are neglected altogether. Taylor-expanding 
the action around the reference path and performing the Gaussian functional integrations 
term by term,  a variant of the systematic eikonal expansion of the scattering amplitude 
derived by Wallace \cite{Wall} is obtained 
(higher orders have been calculated in ref. \cite{Sark}). This is very promising
for applications to many-body scattering as the eikonal
approximation is the basis of Glauber's very successful theory of high-energy scattering from
composite targets \cite{Glau,EiKo}.

Given this affinity to a geometric description of high-energy scattering
and the success of Feynman's treatment of the polaron problem 
it seems interesting to study how a variational approach to potential scattering performs 
in this framework. With a simple (linear)
{\it ansatz} for the trial action this has been investigated in ref. \cite{Carr} where 
it was found that the classical trajectory  -- and not a straight-line path as in the eikonal
approximation -- determines the scattering dynamics \footnote{Corrections to the straight-line 
trajectory also have
turned out to be important in  heavy-ion collisions \cite{Shuk}.}.
Numerically, very promising results in potential scattering have been obtained in cases where 
the eikonal expansion fails. 

It is the purpose of the present work to generalize this work by allowing for 
the most general {\it quadratic + linear} trial action. 
One may expect that the additional quadratic term describes the wave-spreading
characteristic for the exact quantum theory thus leading to a much better description of 
the scattering process.

To be in agreement with the high-energy eikonal expansion such an {\it ansatz} 
must allow for anisotropic terms which have already been shown to improve a variational 
calculation in a scalar field theory \cite{WC7}. Since the Feynman-Jensen variational principle is
the first term of a cumulant expansion it is also possible to calculate systematic corrections. 
We do it here by evaluating the second cumulant which is similar as in the polaron problem 
\cite{polaron cum2,LuRo} but also more challenging as we have to deal with 
the complex scattering amplitude.

The paper is organized as follows: In sect. 2 we present the essentials of the path-integral 
representations of the  ${\cal T}$-matrix in potential scattering so that we can apply
the Feynman-Jansen variational principle in this setting. Sect. 3 contains the variational 
{\it ansatz} and derives the ensuing variational equations. Some properties of the solutions and 
special cases are then discussed and the correction by the second cumulant is given.
Sect. 4 then presents our numerical results for high-energy 
scattering from a Gaussian potential and comparison with exact partial-wave calculation of the
scattering amplitude as well with other approximations discussed in the literature.
The work concludes with a summary and outlook for further work and application.
Most of the technicalities as calculation of various path-integral averages and numerical details
% 2 --> 4 appendices   15. 9.
are collected in four appendices.
 
%%%%%%%%%%%%%%%%%%%%%%%%%%%%%%%%%%%%%%%%%%%%%%%%%%%%%%%%%%%%%%%%%%%%%%%%%%%%%%
%%% ********************** PATH INTEGRALS FOR T-MATRIX **********************
%%%%%%%%%%%%%%%%%%%%%%%%%%%%%%%%%%%%%%%%%%%%%%%%%%%%%%%%%%%%%%%%%%%%%%%%%%%%%

\section{Path-integral representations of the scattering amplitude}
\setcounter{equation}{0}

Recently two variants of a path-integral representation for the ${\cal T}$-matrix in potential 
scattering have been given \cite{Rose} in the form
\be
{\cal T}_{i \to f} \E i \frac{K}{m}
\> \int d^2 b \> e^{- i \fq \cdot \fb } \> \Bigl [ \, S(\fb) - 1 \, \Bigr ] \> ,
\label{PI for T}
\ee
where 
\bea
\fK \EA \frac{1}{2} \left ( \fk_i + \fk_f \right ) \> \> , \> \> K \equiv |\fK| = k 
\cos \frac{\theta}{2}
\\
\fq \EA \fk_f - \fk_i \> \> , \> \> q \equiv |\fq| = 2 k \sin \frac{\theta}{2}
\eea
are the mean momentum  and momentum transfer, respectively. $ k^2/(2m) $ is the scattering energy 
(we set $\hbar = 1$)
and 
$\theta$ the scattering angle. For lack of a better nomenclature we will call $S(\fb)$ the 
``impact-parameter ${\cal S}$-matrix'' although eq. (\ref{PI for T}) is {\it not} a strict  
impact-parameter representation of the scattering amplitude. This is because of the 
dependence of $S(\fb)$ on 
additional kinematic variables like $\fK$ or $\fq$ which we do not show explicitly and the 
angle-dependent factor $ K = k \cos(\theta/2)$ in front of the impact-parameter integral: in a 
genuine impact-parameter representation all dependence on the scattering angle $\theta$ 
should only reside in the factor $\exp ( - i \fq \cdot \fb )$ \cite{impact}.

The main features of these representations are functional integration  over velocities without 
boundary conditions and the use of ``phantom'' 
degrees of freedom to get rid of explicit phases which would diverge in the limit of large scattering 
times. Two versions exist which are distinguished by the reference path along which the particle 
dominantly travels and the dimensionality $d$ of the ``anti-velocity'' $\fw(t)$ which is needed 
to achieve the cancellation~\footnote{The path 
integrals are normalized such that $S(\fb) \equiv 1 $ for zero potential. Our notation indicates that 
$\chi$ is a function of $\fb$ but a 
functional of $\fv(t)$ and $\fw(t)$. Similarly for $\fx_{\rm quant}$.} :
\begin{subequations}
\bea
&& \!S (\fb) \E \int {\cal D}^3 v \, {\cal D}^d w \> \exp \left \{ \,
i \int\limits_{-\infty}^{+\infty} dt \,  \frac{m}{2} \left [ \fv^2(t) -\fw^2(t) \right ] \, 
\right \} \non 
\label{PI for S(b)}
&& \hspace*{4.2cm} \times \exp \Bigl \{ \, i \, \chi(\fb,\fv,\fw] \, \Bigr\} \\
&& \!\! \chi(\fb,\fv,\fw] \E - \int\limits_{-\infty}^{+\infty} dt \, V \left ( \, \fx_{\rm ref}(t) + 
\fx_{\rm quant}(t,\fv,\fw] \, \right ) .
\eea
\end{subequations}
In the first case the reference path is a straight-line path along the mean momentum
\be
\fx_{\rm ref}^{(d=3)} (t) \E \fb + \frac{\fK}{m} t 
\ee
and the quantum fluctuations are given by
\be
\fx_{\rm quant}^{(d=3)}(t,\fv,\fw] \E \fx_v(t) - \fx_w(0)
\label{quant 33}
\ee
where
\be
\fx_v(t) \E \frac{1}{2} \int_{-\infty}^{+\infty} dt' \> {\rm sgn}(t-t') \, \fv(t') \> , \> \> 
\dot \fx_v(t) \E \fv(t) 
\ee
and $ \> {\rm sgn}(x) = 2 \Theta(x) - 1 \> $ is the sign-function.
We will call that the ``aikonal'' representation because it gives rise to the eikonal approximation of
Abarbanel \& Itzykson (AI) \cite{AbIt} if the quantum fluctuations are neglected altogether.

In the second case the anti-velocity is only 1-dimensional and the reference path is a {\it ray} along 
the initial
momentum for $ t < 0 $ and along the final momentum for $t > 0 $
\bea
\fx_{\rm ref}^{(d=1)} (t) \EA \fb + \left [ \, \hat \fk_i \, \Theta(-t) +  \hat \fk_f \, \Theta(t) \, 
\right ] \, \frac{k}{m} t \non
\EA \fb + \frac{\fK}{m} \, t + \frac{\fq}{2m} \, |t|
\eea
and the quantum fluctuations are given by
\be
\fx_{\rm quant}^{(d=1)}(t,\fv,w] \E \fx_v(t) - \fx_{\perp \, v}(0) - \hat \fK \, x_{\parallel \, w}(0) \> .
\label{quant 31}
\ee
We will call that the ``ray'' representation in the following.

For further details we refer to ref. \cite{Rose}. Here we just note that systematic eikonal-like
expansions can be obtained by Taylor-expanding the potential around the reference path
\bea
\chi(\fb,\fv,\fw]  \EA - \int_{-\infty}^{+\infty} dt \> \Bigl [ \, V(\fx_{\rm ref}) + 
\fx_{\rm quant} \cdot \nabla  V(\fx_{\rm ref})  \non
&& \hspace*{-1cm} + \frac{1}{2} \left ( \fx_{\rm quant} \right )_i \, \left ( \fx_{\rm quant}
 \right )_j \, \partial_i \partial_j  V(\fx_{\rm ref}) + \ldots \Bigr ] 
\label{HE expansion}
\eea
and performing successively the functional (Gaussian) integrations over velocity and anti-velocity. 
This is because a 
simple scaling argument shows that each quantum fluctuation $ \fx_{\rm quant} $ is suppressed by
a power of $ 1/\sqrt{K} $ in the case of the ``aikonal'' representation or $ 1/\sqrt{k}$ in the
``ray'' representation. Therefore at high energy/small scattering angle 
the geometrical (classical) picture of scattering is dominant.

\vspace{0.5cm}

%%%%%%%%%%%%%%%%%%%%%%%%%%%%%%%%%%%%%%%%%%%%%%%%%%%%%%%%%%%%%%%%%%%%%%%
%%%% ************ VARIATIONAL CALCULATION *****************************
%%%%%%%%%%%%%%%%%%%%%%%%%%%%%%%%%%%%%%%%%%%%%%%%%%%%%%%%%%%%%%%%%%%%%%%

\section{Variational calculation}
\setcounter{equation}{0}

\subsection{The Feynman-Jensen variational principle}
Being highly nonlinear in velocity and anti-velocity variables the path integrals in 
eq. (\ref{PI for S(b)}) cannot be performed analytically in general \footnote{It would be 
interesting to derive the few exact expressions 
for scattering amplitudes of a local potential which are available in the quantum-mechanical literature, 
e.g. for the Coulomb potential.}. Numerical methods or approximations are then necessary. For weak 
potentials, for example, one may expand in powers of the potential and
it has been shown that the Born series for the ${\cal T}$-matrix is obtained in all orders \cite{Rose}.
At high energy and small scattering angles eikonal approximations are useful.
Because a 
variational approach neither requires weak interaction  nor high energy, forward scattering it is 
widely used
in atomic and molecular physics, mostly in the form of Kohn's variational principle \cite{Kohn}. 
Schwinger's functional
is also stationary against variation of trial wave {\it functions} \cite{LevSchw}
but more difficult to use in practice.

The path-integral representation (\ref{PI for S(b)}) immediately suggests another variational 
approximation for the {\it action}
\bea
{\cal A}[\fv,\fw] \EA \Iint \> \Bigl [ \, \frac{m}{2} \left ( \fv(t)^2 - \fw(t)^2 \right ) \non 
&& \hspace*{0.6cm}  - V \biggl ( \fx_{\rm ref}(t)
+ \fx_{\rm quant}(t,\fv,\fw] \biggr ) \, \Bigr ] \, ,
\eea
{\it viz.} the Feynman-Jensen variational principle. For a positive weight function this principle employs 
the convexity of $\exp(-x)$ to obtain the inequality 
\be
\la e^{-\Delta {\cal A}} \rat \> \ge \> e^{- \la \Delta {\cal A} \rat }
\ee
but in real time (in which scattering occurs) one only has stationarity:
\bea
S(\fb) \EA \int  {\cal D}^3 v {\cal D}^d w \, e^{i {\cal A}_t} \, \frac{\int{\cal D}^3 v 
{\cal D}^d w
\,\exp \left ( i {\cal A}_t + i {\cal A} - i {\cal A}_t \right ) }{  \int{\cal D}^3 v {\cal D}^d w \, 
\exp \left ( i {\cal A}_t \right ) } \non
&\stackrel{\rm stat}{\sim}& S_t(\fb) \cdot e^{ i \left < \Delta {\cal A}  \right >_t } 
\>, \> \>  \Delta {\cal A} \> \equiv \> {\cal A} - {\cal A}_t 
\eea
where
\be
\left < \Delta {\cal A} \right >_t \Def \frac{ \int {\cal D}^3 v {\cal D}^d w \> \, 
\Delta {\cal A} \>  \, \exp \left ( i {\cal A}_t \right ) }{ \int {\cal D}^3 v {\cal D}^d w \>  
\exp \left ( i {\cal A}_t \right )}
\label{def average}
\ee
is the average of the difference between the full action ${\cal A}$ and the trial action  ${\cal A}_t$
weighted with the oscillating factor $\exp (i {\cal A}_t )$. Note that both Kohn's and 
Schwinger's variational principle also are only stationary when applied to the full ${\cal T}$-matrix. 
This seems inevitable when trying to estimate a complex quantity by variational means; only for real 
quantities, like a scattering length, a minimum principle is available (see Chapt. 11.3.4 in ref.
\cite{Newt}).

\subsection{Variational ansatz and equations}

As usual in variational calculations the outcome crucially depends on the test functions/actions which
must lead to expressions which may be evaluated safely.
Unfortunately in the path-integral formalism one is restricted to trial actions which are at most 
{\it quadratic} in the dynamical variables so that the various
path integrals and averages can be worked out analytically. This restricts somewhat the utility of 
this approach which -- on the other hand -- is so general that it can be applied not only for 
the scattering of a single particle in quantum mechanics but also in the many-body case or even in 
field theory. In the present case it means that our trial action ${\cal A}_t$ may contain linear and 
at most quadratic terms in the velocity {\it and} the anti-velocity. So we may take 
\bea
{\cal A}_t[\fv,\fw] \EA \int_{-\infty}^{+\infty} \! dt \, dt' \> \frac{m}{2} \, \Biggl [  \> \> \fv(t) \, 
A_{vv}(t,t') \, \fv(t') \non
&& \hspace*{2.2cm} + \, \fw(t) \, A_{ww}(t,t') \, \fw(t') \non
&& \hspace*{2.2cm} +  \, \fv(t) \, A_{vw}(t,t') \, \fw(t') \non
&& \hspace*{2.2cm} +  \, \fw(t) \, A_{wv}(t,t') \, \fv(t') \>  \> \Biggr ] \non
&&  \hspace*{-0.3cm} 
+ \,  \Iint \> \Bigl [ \, {\bf B}_v(t) \cdot \fv(t) + {\bf B}_w(t) \cdot \fw(t) \, \Bigr ] \quad
\eea
since constant terms cancel out in the Feynman-Jensen variational principle. 
Special cases are the free action 
and the {\it ansatz} studied in ref. \cite{Carr} where only the ${\bf B}$-terms were varied while
the quadratic part was left as in the free case.

At this stage it is useful to have a look at the high-energy expansion in eq. (\ref{HE expansion}): 
It tells us that the functions $ A_{vv}, A_{ww}, A_{vw}, A_{wv} $ have to be {\it anisotropic} in order 
to describe the high-energy expansion up to and including terms of order $ 1/K^2$ (in the ``aikonal'' 
representation) or $1/k^2$ in the ``ray'' representation. Thus the various variational functions 
$A_{vv}(t,t') \dots $ actually are $ 3 \times 3$ matrices in cartesian space if the most general
quadratic trial action is considered. Similar anisotropic trial actions have already been considered 
in a 
variational description of world-line scalar field theory \cite{WC7} and shown to give considerable 
improvement. A slight complication here is the presence of the anti-velocity which, however, can be 
elegantly handled by grouping it together with the velocity to form a $(3+d)$-dimensional ``super-vector''
\be
\V^{(3-3)} \E \left ( \begin{array}{c} \fv \\ \fw \end{array} \right ) \> , \quad \mathrm{or} \quad
\V^{(3-1)} \E \left ( \begin{array}{c} \fv \\ \fw_\parallel \end{array} \right ) 
\ee
so that the trial action can be written succinctly as
\be
{\A}_t \E \frac{m}{2} \> \V \cdot \A \, \V + \B \cdot \V \> \> \> .
\label{A_t compact}
\ee
Here $ \B \> $ is a $(3+d)$-dimensional vector made up of the variational functions ${\bf B}(t)$
\be
\B \E \left ( \begin{array}{c} \fB_v \\ \fB_w \end{array} \right ) 
\ee
and  $ \A \> $ a $(3+d) \times (3+d)$-dimensional symmetric matrix formed by the variational $3 \times 3$ 
matrix functions $A(t,t')$
\be
\A \E \left ( \begin{array}{cc}
              A_{vv} & A_{vw} \\
              A_{wv} & A_{ww} \end{array} \right ) \> .
\ee
In eq. (\ref{A_t compact}) we employ a nomenclature where also the integration over continuous times is
treated like a summation over identical indices. See app. \ref{app: averages} 
for a more detailed account of our conventions.
This allows to evaluate the various path integrals and averages in 
an efficient way 
% new 3. 9. 
and to derive the variational equations easily as detailed in app. \ref{app: var eqs}.
\vspace{0.2cm} 

Here we just collect the final results: after variation the stationary value of the
impact-parameter ${\cal S}$-matrix reads
\be
S_{\rm var}(\fb) \E \exp \left [ \, i \left ( X_0 + X_1 \right ) - \Omega \, \right ] 
\label{var S}
\ee
where 
\bea
X_0 &\equiv& \la \chi \rat \E - \Iint \> V_{\sigma(t)} \left ( \vecrho(t) \right ) 
\label{X0 var}\\
X_1 \EA \frac{1}{2} \, \Iint dt' \> 
\partial_i V_\sigma(t)(\vecrho(t)) \, \left (\Sigma_0\right)_{ij}(t,t') \non
&& \times \partial_j V_{\sigma(t')}(\vecrho(t')) 
\equiv \frac{1}{2} \, J_{\sigma}^T \, \Sigma_0 \, J_{\sigma}  
\label{X1 var} 
\eea 
are ``phases'' and 
\be
\Omega \E \frac{1}{2} {\rm Tr} \, \bigl [ \, - {\rm Ln} \left (1 + \Sigma \, H_{\sigma} \right ) +
\Sigma H_{\sigma} \, \bigr ] \> .
\label{Om var}
\ee
is the ``fluctuation term'' (or an imaginary phase) which arises
from the functional determinant due to the quadratic term in our 
{\it ansatz}. ``Tr'' denotes the trace both in continuous and discrete variables
and ``T'' the transpose of a vector or a matrix.
% new 10. 3. 10
Actually, the terminology used above is a little bit misleading as it turns out that all these 
quantities become complex.
This is because they are not determined by the real potential and its derivatives but by 
its Gaussian transforms with a {\it complex} width $\sigma(t)$ (see below).
This entails that different branches of the multi-valued logarithmic function
``Ln'' may be needed depending on the appropriate sign of the square root of 
the {\it complex} functional determinant. These subtleties of the  complex Gaussian integral 
which require a  ``branch tracking'' are discussed in app. \ref{app: Gauss int}.

The Gaussian transform of the potential is most conveniently defined in momentum space as
% missing tilde inserted 11. 9.
\bea
\tilde V_{\sigma(t)}(\fp) &:=& \tilde V(\fp) \, \exp \left [ - \frac{1}{2} \fp^T \sigma(t) \fp \right ] \non
\EA \tilde V(\fp) \, \exp \left [\,  - \frac{1}{2} \, p_i \sigma_{ij}(t) p_j \, \right ] \> ,
\label{def Vsigma}
\eea
and we can form the Jacobian (the vector of derivatives)
\be
\left ( J_{\sigma} \right )_i \Def \partial_i \, V_{\sigma} 
\ee 
and the Hessian (the matrix of second derivatives) 
\be
\left ( H_{\sigma} \right )_{ij} \Def \partial_i \partial_j \, V_{\sigma} \> .
\label{def Hesse}
\ee
They are functions of a trajectory
$\vecrho_{\rm var}(t)$ (we will omit the subscript ``var'' in the following) 
which is determined by the following equation of motion
\be
\vecrho_{\rm var}(t) \E \fx_{\rm ref}(t) + \Iint' \> \Sigma_0(t,t') \,
{\bf J}_{\sigma(t')} \left ( \vecrho_{\rm var}(t') \right  ) \> .
\label{var rho}
\ee
Here we encounter the $ (3 \times 3)$-matrix $\Sigma_0$ which in the
``aikonal'' representation takes the form
\be
\Sigma_0^{(3-3)}(t,t') \E - \frac{1}{2m} \, |t-t'| \,  \left ( \begin{array}{ccc}
                                                              1 & 0 & 0 \\
                                                              0 & 1& 0 \\
                                                              0 & 0 & 1  \end{array} \right )
\label{Sigma0 33}
\ee
whereas in the ``ray'' representation it reads
\begin{subequations}
\be
\Sigma_0^{(3-1)}(t,t') \E - \frac{1}{2m} \,   \left ( \begin{array}{ccc}
                                                              d(t,t')  & 0 & 0 \\
                                                              0 & d(t,t') & 0 \\
                                                              0 & 0 & |t-t'|  \end{array} \right )
\label{Sigma0 31}
\ee
with
\be
d(t,t') \E |t-t'| - |t| - |t'|
\ee
\end{subequations}
if the mean momentum $\fK$ is chosen to be along the 3-direction. Finally, it is found that
the time-dependent width of the Gaussian transform
\be
\sigma(t) \E i \, \Sigma(t,t)
\label{sig from Sigma}
\ee
is proportional to the diagonal part of a matrix $\Sigma$ 
which fulfills a Lippmann-Schwinger-like equation
\be
\Sigma \E \Sigma_0 + \Sigma_0 \, H_{\sigma} \, \Sigma \> .
\label{LS eq}
\ee
In contrast to the familiar Lippmann-Schwinger equation in scattering theory 
this is a {\it nonlinear} equation since 
the Gaussian width of the potential and its Hessian depends on the diagonal part of $\Sigma$.
%This complication is somehow compensated by the fact that we only have to deal with a (matrix) 
%equation in
%1-dimensional time co-ordinates and not with a scalar equation in 3 spatial co-ordinates or momenta.
Note that eq. (\ref{LS eq}) holds 
for both representations and that all dependence on the phantom degrees of freedom has now disappeared.
In particular, no divergences are encountered for large scattering times -- 
the job for which the anti-velocity was introduced in the beginning. 
We find it also remarkable that in the final outcome
the variational functions $\A, \B$ whose physical interpretation is not immediately evident are  
replaced by the trajectory $\vecrho$ and the Green function $\Sigma$ which have a clear classical and 
quantum-mechanical meaning.
\vspace{0.2cm}

\noindent
Eqs. (\ref{var rho}), (\ref{LS eq}) and (\ref{sig from Sigma}) 
form a system of coupled variational equations which have to be solved. Once that is done we may 
insert the solutions into the variational functional to get the impact-parameter ${\cal S}$-matrix 
(\ref{var S}).
%\ett

\subsection{Correction by the second cumulant}

It is possible to calculate systematic corrections to the variational result 
(or improving the variational principle) by realizing that the 
Feynman-Jensen variational principle is the first term of a {\it cumulant expansion}
\be
\left < \, e^{i \Delta {\cal A}} \, \right >_t \E \exp \left [ \, i \lambda_1 + \frac{i^2}{2!} \, 
\lambda_2 + \ldots \right ]
\ee
where 
\be
\lambda_1 \E \bigl < \, \Delta {\cal A} \, \bigr >_t
\ee
is just what enters the Feynman-Jensen variational principle and
\be
\lambda_2 \E  \left < \left (\Delta {\cal A} \right )^2  \right >_t -  \bigl <  \, \Delta {\cal A} 
\bigr >_t^2
\ee
the first correction to it. So the variational functional becomes
\be
S_{\rm var +cum}(\fb) \E \exp \left [ \, i \left (X_0 + X_1 \right ) - \Omega - \frac{1}{2} \, 
\lambda_2 \, \right ] \> .
\label{S(b) var+cum}
\ee
In principle one could vary the full functional with respect to the variational functions/parameters 
but we follow the standard practice to
add the second cumulant as a correction \cite{polaron cum2} 
with the parameters fixed by the Feynman-Jensen variational principle, i. e. the first cumulant.
This allows to use the variational equations for these parameters 
and leads to a considerable simplification of the final expression.

App. \ref{app: cum2} gives the result of evaluating the second cumulant in our case:
\be
\lambda_2 \E  \la \chi^2 \rat - \la \chi \rat^2 - i J_{\sigma}^T \, \Sigma \, J_{\sigma} 
+ \frac{1}{2} \, {\rm Tr} \left ( \Sigma H_{\sigma} \right )^2
\label{lambda_2}
\ee
where
\bea
\la \chi^2 \rat \EA \Iint_1 \, dt_2 \, \int \frac{d^3 p_1 d^3 p_2}{(2 \pi)^6} \> \> 
\tilde V(\fp_1) \, \tilde V(\fp_2) \non
&& \hspace*{-1cm} \times \, \exp \Biggl  \{ \, - \frac{i}{2} \left ( \fp_1^T,\fp_2^T \right ) \, 
\left ( \begin{array}{cc} \Sigma(t_1,t_1) & \Sigma(t_1,t_2) \\
               \Sigma(t_2,t_1) & \Sigma(t_2,t_2) \end{array} \right ) \, \left ( \begin{array}{c} 
                                                                 \fp_1 \\
                                                                 \fp_2 \end{array}  \right ) \non 
  && \hspace*{0.3cm} + \,    i  \, \left ( \fp_1^T,\fp_2^T \right ) \, \left (
                \begin{array}{c} \vecrho(t_1) \\
                                \vecrho(t_2) \end{array} \right ) \, \Biggr \} 
\label{<chi2>}
\eea
involves a double Gaussian transform of the squared potential.
All other quantities have been defined and calculated before in both representations.
It is easy to see that $ \Delta {\cal A} $ is of first (and higher) order in the potential. 
By construction the $n^{\rm th}$ cumulant then contains 
contributions of  $ {\cal O}(V^n)$ and higher. This implies that the second Born approximation to 
the ${\cal T}$-matrix is fully included in our calculation when the second cumulant is added.

\subsection{A special case: The linear ansatz}

A less general variational {\it ansatz} has been made in ref. \cite{Carr} by only allowing the 
linear terms in the trial action to vary whereas the quadratic part was fixed to be the 
free action. In the present nomenclature this amounts to setting $ \A = \sigma_3 $. 
Inspecting the variational solution for $\A$ we see that the results of ref. {\cite{Carr} 
should be recovered by setting the Hessian of the potential 
to zero: $ H_{\sigma} \to 0 $. Indeed, then the solution of the Lippmann-Schwinger eq. (\ref{LS eq}) 
simply is $ \Sigma =  \Sigma_0 $ and the Gaussian width becomes
\be
\sigma_{ij}(t) \Bigr |_{\rm linear} \E - \frac{i |t|}{m} \,  
\delta_{ij} \> \left \{ \begin{array}{ll}
                                          0 & \quad {\rm ``aikonal''}\\
                                     \left ( 1 - \delta_{i3} \right ) &  \quad {\rm ``ray''} \> .
                                              \end{array} \right . 
\label{width for B ansatz}
\ee
Thus in the ``aikonal'' case (with a 3-dimensional anti-velocity) there is no Gaussian transform of 
the potential and the variational equation for the trajectory simply is
\be
\vecrho(t) \E \fb + \frac{K}{m} t - \frac{1}{2m} \Iint' \> |t - t'| \, \nabla V(\vecrho(t'))
\label{int for rho}
\ee
which, after differentiating twice, is just  Newton's law for the classical motion in  the potential:
\be
\ddot{\vecrho}(t)  \E - \frac{1}{m} \,  \nabla V(\vecrho(t)) \> .
\label{Newton}
\ee
However, the boundary conditions encoded in the integral eq. (\ref{int for rho}) are unusual: using
$ |t-t'| \to |t| - {\rm sgn}(t) \, t' $ for large $|t|$ one finds from eq. (\ref{int for rho})
\bea
\vecrho(t) & \stackrel{ t \to \pm \infty}{\longrightarrow} & \left [ \frac{\fK}{m} \mp  
\, \int\limits_{-\infty}^{+\infty} dt'\, \frac{{\bf J}(t')}{2m} \, \right ] \, t 
+ \fb \pm  \,   \int\limits_{-\infty}^{+\infty} dt' \, t' \, \frac{{\bf J}(t')}{2m} \non
&& \hspace*{0.7cm} +  \, \mbox{terms which vanish for} \, |t| \to \infty \> ,
\label{asy rho}
\eea
%where $ {\bf J(t)}  = \nabla V(\vecrho(t)) $ is the Jacobian of the potential 
so that
\begin{subequations}
\bea
\hspace*{-1cm} \lim_{T \to \infty} \left \{ \vecrho(T) + \vecrho(-T) - T \left [  \, \dot{\vecrho}(T) - 
\dot{\vecrho}(-T) \, \right ] \right \} \! \EA \! 2 \, \fb 
\label{bc 33 a}\\
\lim_{T \to \infty} \left \{ \,  \dot{\vecrho}(T) + \dot{\vecrho}(-T) \, \right \} \! \EA \! 2 \, 
\frac{\fK}{m} .
\label{bc 33 b}
\eea
\end{subequations}
It is the great advantage of the integration over velocities that the boundary conditions are exactly 
built in and we do not have to worry about them during (approximate) evaluation of the path integral.
Note that the boundary conditions (\ref{bc 33 a})
and (\ref{bc 33 b}) are just those needed to convert an integral equation 
with the kernel $ \> |t-t'| \> $ into a differential equation \cite{IntEq}. 

For the linear {\it ansatz} in the ``aikonal'' representation the trajectories are real
and consequently the phases $X_0$ and $X_1$ are also real; an imaginary phase, i.e. a real part in 
$ \ln S(\fb)  $ only develops due the second cumulant. It is easily seen that the 
general expression (\ref{<chi2>}) for $\lambda_2$ reduces to the eq. (4.84) in ref. \cite{Carr}, 
if $ \Sigma $ is replaced by $ \Sigma_0^{(3-3)} $.
As usual one finds from Newton's equation (\ref{Newton}) the conservation law
$ \> \> m \dot{\vecrho}^2(t)/2 +  V(\vecrho(t)) = {\cal E} \> $ = const.
Inserting the asymptotic behaviour 
(\ref{asy rho}) and using the symmetry of the solutions under $ t \to -t $ which forces 
$ \Iint' \, {\bf J(t')} $ to be perpendicular to the mean momentum $\fK$, the conserved quantity is 
\be
{\cal E}  \E \frac{\fK^2}{2m} + \frac{1}{8m} \, \left ( \Iint' \> {\bf J(t')} \right )^2 
\ee
which means that this ``energy'' depends on the impact parameter $\fb$.
This is because $S(\fb) = S(\fb,\fK)$ does not contain any information on the actual scattering energy
\be
\frac{k^2}{2m} \E  \frac{\fK^2}{2m} + \frac{\fq^2}{8m} \> .
\ee
However the variational principle is ``clever'' enough to mimic the missing term as good as possible:
\be
\fq \> \simeq \> -  \Iint' \> \nabla_b V(\vecrho(t'))
\ee
is what one would obtain if the impact-parameter integral over $\exp( i X_0 ) $ is evaluated 
by a saddle-point approximation.

Things are slightly different in the ``ray'' representation: First, the potential is replaced by its
Gaussian transform with an anisotropic width given in eq. (\ref{width for B ansatz}) which renders it 
complex from the very beginning. Second, the anisotropic kernel (\ref{Sigma0 31}) leads 
to an equation of 
motion where the particle experiences a kick at $t = 0$ and the ``energy'' of that motion is not 
conserved anymore during the scattering process but only asymptotically (see fig. 1 in 
ref. \cite{Carr}). 
Complex trajectories are frequently encountered in semi-classical approximations of quantum motion 
in regions of space which are forbidden classically. In our variational approximations, however, the 
main reason for this behaviour (which shows up even above any potential hill) 
is unitarity: $S(\fb)$ cannot be unimodular \cite{Wall}. 

We do not dwell on several other interesting properties of our variational approximation with 
a linear {\it ansatz} but refer to ref. \cite{Carr} for more details.
                                                                 
\subsection{High-energy expansion}

By construction the variational solution of the quadratic + linear variational 
{\it ansatz} (\ref{A_t compact}) must contain the high-energy expansion (\ref{HE expansion}). 
To show how that comes out we will take here for simplicity the  
``aikonal'' representation as the algebra in the ``ray'' representation is more involved.

First, we switch to distances instead of times by defining
\be
z \Def \frac{K}{m} \, t \> \> \> \Rightarrow \> \> t \E \frac{m}{K} \, z  \> .
\label{def z}
\ee
We then see that 
\be
\Sigma_0^{(3-3)}(z,z') \E - \frac{1}{2K} \, \left | z - z' \right| \, 
\ee
is suppressed for large energies and forward scattering angles (recall $  K = k \cos(\theta/2) $)
since the arguments $ z,z'$ are 
bounded by the range  of the potential. 
We thus may expand in powers of $\Sigma_0$ and obtain
from the Lippmann-Schwinger-like eq. (\ref{LS eq}) 
\be
\Sigma \E \Sigma_0 + \Sigma_0 \, H_{\sigma} \, \Sigma_0 + \ldots \> \> .
\ee
Eq. (\ref{sig from Sigma}) then gives for the Gaussian width 
\be
\sigma_{ij}(z) \E 
\frac{i}{(2K)^2} \, \int_{-\infty}^{+\infty} dz' \> \left | z - z' \right |^2 \, \frac{m}{K} 
\left ( H_{\sigma(z')} \right )_{ij}(z') \> + \> \ldots
\> ,
\ee
as the first-order term vanishes in the ``aikonal'' representation (this is 
not the case in the ``ray'' representation~!). Therefore 
the Gaussian width becomes also small under these kinematical conditions (note that $m/K$ doesn't need 
to be small) so that $H_{\sigma}$ may be replaced by $H$ up to order $1/K^2$.
From the definition (\ref{def Vsigma}) of the Gaussian transform we find
\be
V_{\sigma(z)}(\vecrho(z)) \E V(\vecrho(z)) + \frac{1}{2} \, \sigma_{ij}(z) H_{ij}(\vecrho(z)) + \ldots
\label{HE Vsigma}
\ee
and iterating the equation of motion (\ref{var rho}) one obtains
% changed 25. 8.
\be
\vecrho \E \fx_{\rm ref} + \Sigma_0 \, {\bf J}  + \Sigma_0 \, H  \Sigma_0 \,  {\bf J}  + \ldots \> .
\label{HE rho}
\ee
Here and in the following the potential, the Jacobian and the Hessian 
are all evaluated with the  straight-line reference path
$  \fx_{\rm ref}(z) = \fb + \hat K z $ .
% end change
Inserting these expressions 
into eqs. (\ref{X0 var}) - (\ref{Om var}), and expanding up to order $ 1/K^2$ we get in our short-hand notation
\bea
&& \hspace*{-0.5cm} X_0 \to  \chi_{AI}^{(0)} - J^T \! \Sigma_0  J - \frac{3}{2}  J^T  \! \Sigma_0 H \Sigma_0 J 
- \frac{i}{2}  {\rm Tr} \left ( \Sigma_0 H \right )^2 \quad \\
&&  \hspace*{-0.5cm} X_1  \to   \hspace*{0.8cm} \frac{1}{2} J^T  \Sigma_0  J  +   
J^T  \Sigma_0  H  \Sigma_0  J   \quad \\ 
&&  \hspace*{-0.5cm} \Omega \to \hspace*{5.1cm} \frac{1}{4}   {\rm Tr} \left ( \Sigma_0 H \right )^2 \quad
\label{Omega HE}
\eea
where
\be
 \chi_{AI}^{(0)} \E - \Iint \> V(\fx_{\rm ref}(t)) \E - \frac{m}{K} \int_{-\infty}^{+\infty} dz \> 
V \left (\fb + \hat K z \right )
\ee
is the standard eikonal phase from Abarbanel \& Itzykson \cite{AbIt}. We see that
% new 25. 8.
\be
X_0 + X_1  \To   \chi_{AI}^{(0)} +  \chi_{AI}^{(1)} + \chi_{AI}^{(2)} + 2 i \, \omega_{AI}^{(2)} + 
{\cal O} \left ( \frac{1}{K^3} \right )
\label{X0+X1 HE}
\ee
where in our condensed notation
\bea
\chi_{AI}^{(1)}  \EA - \frac{1}{2} J^T \, \Sigma_0 \, J \> \> , \> \> \> \chi_{AI}^{(2)} \E  
-\frac{1}{2}  J^T \, \Sigma_0 \, H \, \Sigma_0 \, J \\
\omega_{AI}^{(2)} \EA - \frac{1}{4} \, {\rm Tr} \left ( \Sigma_0 H \right )^2 
\eea
are exactly the first- and second-order corrections
to the leading ``aikonal'' result as given in eqs. (4.97), (4.103) and (4.110) of ref. \cite{Carr}.
Note that the contribution (\ref{Omega HE}) from the fluctuation term 
brings the imaginary part into full
agreement \footnote{To show that one has to convert to an impact-parameter representation 
of the scattering matrix for which the $K = k \cos(\theta/2)$-factors in the ``aikonal''
representation give additional contributions. As they are the same as worked out in ref. \cite{Carr} 
we do not have to consider them here.}
with what is needed for unitarity in the high-energy limit:
\be
{\rm Im} X_0 + \Omega \to 2 \omega_{AI}^{(2)}  - \omega_{AI}^{(2)} + 
{\cal O} \left ( \frac{1}{K^3} \right ) = \omega_{AI}^{(2)} 
+ {\cal O} \left ( \frac{1}{K^3} \right ) .
\label{Im+Om HE}
\ee

How is that high-energy behaviour changed by including the second cumulant? To answer this question we
consider eq. (\ref{<chi2>}) and decompose 
\bea
\left ( \begin{array}{cc} \Sigma(t_1,t_1) & \Sigma(t_1,t_2) \\
               \Sigma(t_2,t_1) & \Sigma(t_2,t_2) \end{array} \right ) 
       \EA
\frac{1}{i} \, \left ( \begin{array}{cc} \sigma(t_1) & 0 \\
               0 & \sigma(t_2) \end{array} \right ) \non
&& \hspace*{-0.5cm} +  \, \left ( \begin{array}{cc} 0 & \Sigma(t_1,t_2)\\
               \Sigma(t_2,t_1) & 0 \end{array} \right )  
\eea
into a diagonal and a nondiagonal part. The diagonal part is seen to convert the
potentials into Gaussian 
transformed potentials while the nondiagonal part gives new contributions when the exponential
is expanded:
\bea
\left < \chi^2 \right > \EA \left < \chi \right >^2 \, + \, i J^T \Sigma J \, + \, 
\frac{i^2}{2} {\rm Tr} \left ( H_{\sigma} \Sigma H_{\sigma} \Sigma \right ) \non
&& + \frac{i^3}{6} \, \int_{-\infty}^{+\infty} dt_1 dt_2 \, 
\left ( \partial_{2i} \Sigma(t_1,t_2)_{ij}
\partial_{1j} \right )^3 \non
&& \hspace*{1.5cm} \times \, V_{\sigma(t_1)}(\vecrho(t_1))  V_{\sigma(t_2)}(\vecrho(t_2)) + \ldots \> \> .
\eea
It is seen that the the first 3 terms on the r.h.s. 
are completely cancelled when forming the second cumulant in  
 eq. (\ref{lambda_2}) and that in the high-energy limit the leading contribution comes from the 
last term involving 3 powers of the matrix $\Sigma$.
In that limit we may safely replace it by the zeroth-order matrix
$\Sigma_0$ so that the second cumulant
gives an  imaginary (phase) correction at  ${\cal O}(1/K^3)$. It is interesting that 
in the linear approximation the last term
in eq. (\ref{lambda_2}) is absent and that the contribution $-\lambda_2/2$  
from the second cumulant then correctly supplies the real 
term $\omega_{AI}^{(2)}$.
In contrast, the more general, anisotropic 
quadratic + linear trial action already gives variational results correct up to and 
including  ${\cal O}(1/K^2)$ as expected from the high-energy expansion. Similar results hold in 
the ``ray'' representation except that here 
the expansion parameter is the wave number $k$ and not the mean momentum $ K = k \cos(\theta/2)$.

% new  23. 9. 
\subsection{A Feynman-Hellman theorem}
\label{sec: FH}
Some further properties of the variational approximation for the impact-parameter ${\cal S}$-matrix 
(\ref{var S})
can be obtained by employing the variational equations in a particular way. This has 
already been used in a variational approximation to the relativistic bound-state problem \cite{BaStRo}
and is just a variant of the well-known Feynman-Hellmann theorem in quantum mechanics (see, e.g. refs. 
\cite{Fey-Hell1,Fey-Hell2}).

Suppose there is some parameter
$\lambda$ in the potential or in the reference path. As the variational functional also depends on it 
implicitly via the variational parameters/functions $\A_{\rm var}$ and $\B_{\rm var} $ one may perform the 
differentiation with respect to this parameter 
by means of the chain rule \footnote{Here it is important that only $X_0$ is assumed to depend explicitly 
on $\lambda$ whereas $X_1, \Omega$ do not. See eqs. (\ref{def X0}), (\ref{def X1}), (\ref{def Omega}) 
in app.~\ref{app: var eqs}. We use the short-hand notation of app.~\ref{app: averages}.}
\bea
\frac{\partial \ln S_{\rm var}}{\partial \lambda} \EA 
\frac{\partial \ln S_{\rm var}}{\partial \A^{\rm var}_{AB}} \, 
\frac{\partial \A^{\rm var}_{AB}}{\partial \lambda} +  \frac{\partial \ln S_{\rm var}}
{\partial \B^{\rm var}_A} \, 
\frac{\partial \B^{\rm var}_A}{\partial \lambda} \non
&& - i \Iint \> \frac{\partial}{\partial \lambda} \, 
V_{\sigma(t)}(\vecrho(t)) \non
\EA  - i \Iint \> \frac{\partial}{\partial \lambda} \, 
V_{\sigma(t)}(\vecrho(t)) \> ,
\eea
since the first two terms vanish identically for the variational solutions $\A_{\rm var}$ and 
$\B_{\rm var}$. This leads to a simple result
although  these solutions may be very complicated functions of the parameter $\lambda$.

As an example take the dependence of the impact-parameter ${\cal S}$-matrix on the mean momentum $K$
in the ``aikonal'' representation. 
The sole explicit dependence of the variational functional resides in the reference path $\fx_{\rm ref}(t) = 
\fb + \fK t/m $. Therefore we immediately have the relation
\be
\frac{\partial \ln S_{\rm var}^{(3-3)}}{\partial K} \E 
- \frac{i}{m} \, \hat \fK \cdot \Iint \> t \> 
\nabla V_{\sigma(t)}(\vecrho(t)) \> .
\label{FH K}
\ee
Similarly one obtains 
\be
\frac{\partial \ln S_{\rm var}}{\partial b} \E - i \, \hat \fb \cdot \Iint \>  
\nabla V_{\sigma(t)}(\vecrho(t)) 
\ee
which also holds for the ``ray'' representation.
Finally the dependence on the strength $V_0$ of the potential
is given by the simple expression
\be
V_0 \,  \frac{\partial \ln S_{\rm var}}{\partial V_0} \E - i \Iint \>  V_{\sigma(t)}(\vecrho(t)) 
\> \equiv \> i X_0 \> .
\label{FH V0}
\ee
These relations may be used to derive or check high-energy or weak-coupling expansions by expanding 
the r.h.s. in powers of the particular parameter and integrating term by term. 

% new 10. 12. 09
For example, using the first two terms of the high-energy expansion (\ref{HE rho}) we have in the 
``aikonal'' representation
\bea
&& \hat \fK \cdot \nabla V_{\sigma(t)}(\vecrho(z)) \E \frac{\partial}{\partial z} \, \Bigl [ \, 
V(\fx_{\rm ref}(z)) - \frac{1}{2 K}  \, \nabla V(\fx_{\rm ref}(z)) \non
&&  \hspace{1.2cm} \cdot \, \int dz' \> |z-z'|  \, 
\nabla V(\fx_{\rm ref}(z')) +  {\cal O} (1/K^2) \, \Bigr ]
\eea
since the Gaussian width does not contribute in this order. Substituting 
this into the Feynman-Hellmann relation (\ref{FH K}) (expressed in the $z$-variable to make 
$ \> \fx_{\rm ref}(z) \> $ independent of $K$)
and integrating with respect to $K$ gives the
correct result $ \> \ln S_{\rm var}^{(3-3)} \E i \chi_{AI}^{(0)} + i \chi_{AI}^{(1)} 
+  {\cal O} (1/K^3) \> $ 
if an integration by parts in $z$ is performed.

In app. \ref{app: test of FH} we use the dependence (\ref{FH V0}) on the coupling strength 
as a nontrivial check for our numerical procedures to solve the variational equations.
% end new 10. 12. 09

\subsection{Semi-classical expansion}

% new 1. 9.
It is of some interest to distinguish the present variational approach from the usual semi-classical
approximation of the path-integral propagator obtained by the stationary phase approximation 
(see, e.g. ref. \cite{Schul}, Chapt. 13 or ref. \cite{Klein}, Chapt. 4).
To do that we have 
to restore the appropriate $\hbar$-factors in the path-integral representation (\ref{PI for S(b)}). This 
is easily done by remembering that the weight in the path integral is always the exponential of 
$ i \, \times$ the action divided by $\hbar$ and that the velocities in the reference path are {\it momenta} 
($ = \hbar \, \times $ wavenumbers) divided by the mass of the particle. To account for that we simply 
have to substitute 
\bea
m & \longrightarrow & \frac{1}{\hbar} \, m \\
V & \longrightarrow &  \frac{1}{\hbar} \, V
\label{V hbar}
\eea
everywhere. From the explicit expressions (\ref{Sigma0 33}) and (\ref{Sigma0 31}) we then see that
\be
\Sigma_0 \E {\cal O} ( \hbar)
\ee
and 
from the Lippmann-Schwinger-like equation (\ref{LS eq}) that
\be
\Sigma \E  {\cal O} ( \hbar) + \mbox{ higher orders in } \> \hbar
\label{Sigma hbar}
\ee
since eq. (\ref{V hbar}) also implies the multiplication of the Jacobian and Hessian by $1/\hbar$.
In other words: the Gaussian width $\sigma(t) \equiv \Sigma(t,t)$ is a pure quantum effect which 
vanishes in the classical approximation. The ``higher orders in $\hbar$ `` in eq. (\ref{Sigma hbar}) 
are generated by the nonlinearity of the Lippmann-Schwinger-like equation 
via the Gaussian transform of the Hessian $H_{\sigma}$. Similarly the Gaussian transform of the 
Jacobian $J_{\sigma}$ in eq. (\ref{var rho}) gives rise to $\hbar$-dependent terms in
the trajectory 
\be
\vecrho_{\rm var}(t) \E \vecrho_{\rm class}(t)  + {\cal O} ( \hbar) + \ldots  \> ,
\ee
where $ \vecrho_{\rm class}(t) $ is the trajectory for zero Gaussian width.
By the same arguments we see from eqs. (\ref{X0 var}), (\ref{X1 var}) that the phases
\be
X_0 , \, X_1 \E {\cal O} \left ( \frac{1}{\hbar} \right ) + {\cal O} ( \hbar^0 ) + \ldots \> ,
\ee
whereas from eq. (\ref{Om var}) one finds that
\be
\Omega \E  {\cal O} ( \hbar^0 ) + \ldots \> \> .
\ee
Finally, applying the scaling laws (\ref{V hbar}) and (\ref{Sigma hbar}) to eqs. (\ref{lambda_2}), 
(\ref{<chi2>} we deduce that
\be
\lambda_2 \E {\cal O} ( \hbar ) + \ldots 
\ee
is a genuine quantum correction.
We thus see that our variational approach is {\it not} equivalent to a semi-classical approximation as
it contains (an infinite number of) $\hbar$-dependent terms. This is mostly due to the quadratic term in our 
general variational {\it ansatz} (\ref{A_t compact})
which gives rise to an interaction-dependent Gaussian width:
the more restricted linear {\it ansatz} studied 
in ref. \cite{Carr} leads to a 
vanishing Gaussian width in the ``aikonal'' (3-3) representation; only in the ``ray'' (3-1) representation
one obtains a nonzero width which generates some quantum corrections.

%%%%%%%%%%%%%%%%%%%%%%%%%%%%%%%%%%%%%%%%%%%%%%%%%%%%%%%%%%%%%%%%%%%%%%%
%%%% ************ NUMERICAL RESULTS **********************************
%%%%%%%%%%%%%%%%%%%%%%%%%%%%%%%%%%%%%%%%%%%%%%%%%%%%%%%%%%%%%%%%%%%%%%

\section{Numerical results}
\setcounter{equation}{0}

%%%%% begin 26. 6. 09
We compare our results with those of a (basically) exact partial-wave 
calculation 
of scattering from a Gaussian potential 
\be
V(r) \E V_0 \, e^{-\alpha r^2} \E V_0  \, e^{- r^2/R^2}
\label{Gauss pot}
\ee
with parameters
\bea
2 m V_0 R^2 \EA - 4 \non
k R \EA 4  \> \> .
\label{param}
\eea
This was considered as a test case for the systematic eikonal expansion 
by Wallace in ref. \cite{Wall} and (with $V_0 = - 41.6$ MeV and 
$R = 1 $ fm) used to describe phenomenologically
the scattering of $\alpha$-particles from  $\alpha$-particles
at an energy of $ 166 $ MeV. We also choose it because for 
this potential a persistent failure of the eikonal expansion  
to describe the scattering at larger angles was observed: the scattering angle
where the eikonal amplitude started to deviate appreciably increased only slightly
when higher-order corrections were included
(see fig. 6 in ref. \cite{Wall}).

Thus this particular potential is a good {\it litmus test} for checking 
any approximate
description of high-energy scattering. However, we do not use
\be
\frac{\Delta \sigma}{\sigma} \Def \frac{d\sigma/d\Omega|_{\rm approx} - 
d\sigma/d\Omega|_{\rm exact}}
{ d\sigma/d\Omega|_{\rm exact}}
\label{def delta sigma}
\ee
as the usual measure to gauge agreement/disagreement with the exact result but
\be
\left | \frac{\Delta f}{f} \right | \Def \left | \frac{f_{\rm approx} - 
f_{\rm exact}}{ f_{\rm exact}} \right | \> .
\label{def delta f}
\ee
Here 
\be
f(\theta) \E - \frac{m}{2 \pi} \, {\cal T}_{i \to f}
\ee
is the scattering amplitude and
\be 
\frac{d \sigma}{d \Omega} \E \left | f(\theta) \right |^2
\ee
the differential cross section.
By construction the quantity in  eq. (\ref{def delta f}) also measures the {\it phase} deviation 
of the scattering amplitude $f_{\rm approx}$ from the exact amplitude. Indeed, if
\be
f_{\rm exact} \E r \, e^{i \phi} \> \> \> \>  {\rm and} \> \> \> \> 
f_{\rm approx} \E \left ( r + \Delta r \right ) e^{i (\phi + \Delta \phi)}
\ee
then one finds to first order in the deviations $\Delta r , \Delta \phi$
\be
\left | \frac{\Delta f}{f} \right | \E \sqrt{ \left ( \Delta \phi \right )^2 
+ \left (  \Delta r / r \right )^2 } 
\ee
and
\be
\frac{\Delta \sigma}{\sigma} \E 2 \frac{\Delta r}{r}
\ee
which is independent of the phase error $\Delta \phi$. As
\be
\left | \frac{\Delta \sigma}{\sigma} \right | \le 2 \, 
\left | \frac{\Delta f}{f} \right |
\ee
one may even encounter cases where $ \Delta \sigma/\sigma = 0 $ although
$\left | \Delta f/f \right| \ne 0$ because the approximate scattering
amplitude has the correct modulus but the wrong phase. 
So the common practice of
comparing just differential cross sections can be misleading.

%%%%% end 26. 6. 09

%%%%%%%%%%%%%%%%%  Table exact ( 29. 6. 09)
%\vspace{-0.1cm}

\begin{table}[htb]
\caption{Exact partial-wave amplitude and differential cross section for scattering from the
Gaussian potential (\ref{Gauss pot}) with $ 2 m V_0 R^2 = - 4 $ at $k R = 4 $. Amplitude and
cross section are given in units of $R$ and $R^2$, respectively, where $R$ is the range of the 
potential. The number in parenthesis indicates 
the power of ten by which the numerical value has to be multiplied, e.g. $  5.6472 \, (-6) \equiv
5.6472 \cdot 10^{-6}$.
}
\label{tab: exact}
\begin{tabular}{rccc} \hline\noalign{\smallskip}
                      &                           &              &    \\
$\theta \> \>  [ \, ^{\circ}\, ] $   & \quad  (Re $f)/ R$  &  (Im $f)/ R$ \quad &  \quad $ 
\frac{d\sigma}{d\Omega}/ R^2$ \hspace{0.5cm} \\
                      &                           &              &    \\
\noalign{\smallskip}\hline\noalign{\smallskip}
  0  \hspace{0.4cm}    &   1.7348 \,(0)  &     3.8758 (-1)    &   3.1596 \,(0) \hspace{0.4cm} \\
  5   \hspace{0.4cm}   &   1.6806 \,(0)  &     3.8172 (-1)   &   2.9702 \,(0) \hspace{0.4cm} \\
 10  \hspace{0.4cm}   &    1.5282 \,(0) &      3.6466 (-1)  &     2.4683 \,(0) \hspace{0.4cm} \\
 15  \hspace{0.4cm}   &    1.3042 \,(0) &      3.3788 (-1) &      1.8151 \,(0) \hspace{0.4cm} \\
 20  \hspace{0.4cm}   &    1.0445 \,(0) &     3.0358 (-1)   &    1.1831 \,(0)  \hspace{0.4cm} \\
 25  \hspace{0.4cm}   &    7.8431 (-1) &     2.6444 (-1)  &    6.8507 (-1)   \hspace{0.4cm} \\
 30  \hspace{0.4cm}   &    5.5114 (-1)&      2.2323 (-1) &     3.5358 (-1) \hspace{0.4cm} \\
 35   \hspace{0.4cm}  &    3.6078 (-1)&        1.8249 (-1)&     1.6347 (-1) \hspace{0.4cm} \\
 40  \hspace{0.4cm}  &     2.1785 (-1)&        1.4435 (-1)&     6.8294 (-2)  \hspace{0.4cm} \\
 45 \hspace{0.4cm}  &      1.1867 (-1)&        1.1032 (-1)&     2.6252 (-2)  \hspace{0.4cm} \\
 50  \hspace{0.4cm} &      5.5103 (-2)  &   8.1274 (-2) &     9.6419 (-3)  \hspace{0.4cm} \\
 55 \hspace{0.4cm}  &      1.7782 (-2) &    5.7530 (-2) &     3.6259 (-3)  \hspace{0.4cm} \\
 60 \hspace{0.4cm}  &     -1.8100 (-3) &       3.8913 (-2) &    1.5175 (-3)  \hspace{0.4cm} \\
 65  \hspace{0.4cm} &     -1.0380 (-2) &        2.4924 (-2) &   7.2893 (-4) \hspace{0.4cm} \\
 70  \hspace{0.4cm} &     -1.2687 (-2) &        1.4870 (-2) &   3.8207 (-4)  \hspace{0.4cm} \\
 75  \hspace{0.4cm} &     -1.1832 (-2) &        7.9883 (-3) &   2.0380 (-4)  \hspace{0.4cm} \\
 80  \hspace{0.4cm} &     -9.6765 (-3)  &      3.5396 (-3) &    1.0616 (-4)  \hspace{0.4cm} \\
 85  \hspace{0.4cm} &     -7.2471 (-3)  &      8.6532 (-4) &    5.3269 (-5)  \hspace{0.4cm} \\
 90  \hspace{0.4cm} &     -5.0442 (-3)  &     -5.8091 (-4) &    2.5782 (-5)  \hspace{0.4cm} \\
 95  \hspace{0.4cm} &     -3.2623 (-3)  &     -1.2260 (-3) &    1.2145 (-5)  \hspace{0.4cm} \\
100  \hspace{0.4cm} &     -1.9307 (-3)  &     -1.3856 (-3) &    5.6472 (-6)  \hspace{0.4cm} \\
%105  \hspace{0.4cm} &     -1.0002 (-3)  &     -1.2799 (-3) &    2.6386 (-6)  \hspace{0.4cm} \\
%110  \hspace{0.4cm} &     -3.925 \, (-4) &      -1.0539 (-3) &    1.265 \, (-6)  \hspace{0.4cm} \\
%115  \hspace{0.4cm} &     -2.615 \, (-5) &      -7.961 \, (-4) &    6.344 \, (-7)  \hspace{0.4cm} \\
%120  \hspace{0.4cm} &      \, 1.707 \, (-4) &     -5.551 \, (-4)  &    3.373 \, (-7)  \hspace{0.4cm} \\
\noalign{\smallskip}\hline

\end{tabular}
\vspace*{1cm}
\end{table}
%%%%%%%%%%%%%%%% end table exact
%\clearpage

\vspace{0.2cm}

Table \ref{tab: exact} gives the values of the exact calculation obtained 
by integrating the Schr\"odinger equation  for each partial wave up to a radius
where the solution could be matched to  a linear combination of free spherical waves. By varying 
this radius as well as the step size for the numerical integration and the number of partial waves
retained we made sure that the numerical values given in this table are at least accurate to the 
number of digits given. We display scattering amplitude and 
differential cross section in units of powers of $R$ since  the range of the potential sets the
length scale.
%\clearpage

\subsection{Variational results} 

We have evaluated the coupled variational equations by an iterative scheme 
on a grid in time (more precisely: in distances which the particle travels
while in the range of the potential)
and then evaluated the phases $X_0, X_1$, the fluctuation term $\Omega$
and the second cumulant $\lambda_2$. Some numerical details are given in app. \ref{app: num details}.

\vspace{0.2cm}

In table \ref{tab: var results} we list the relative deviation of the scattering amplitude 
for the various approximations from
the exact partial-wave result as a function of the scattering angle $\theta$.

%%%%%%%%%%%%%%%%%  Table var results ( 26. 11. 09)
\def\qq{\hspace*{0.55cm}}

\begin{table*}[htb]
\caption{ The relative deviation (in units of $10^{-4}$) of the scattering amplitude from the exact value 
for scattering from the Gaussian potential (\ref{Gauss pot}) with $ 2 m V_0 R^2 = - 4 $ at $k R = 4 $. 
The variational approximations are labeled according to the scheme (\ref{name var approx}).
The numbers in parenthesis give the estimated numerical error of the calculations in units of the last 
digit. Entries are stopped if the deviation exceeds 10 \%.
}
\label{tab: var results}
\bce
\begin{tabular}{rrrrrrrrr} \hline\noalign{\smallskip}
                                   &       &      &       &       &       &       &        &  \\

$\theta \quad  [ \, ^{\circ}\, ] $ & \quad B 33 &  \quad  B 31 & \quad  Bc 33 & \quad  Bc 31 &  \quad AB 33 
& \quad AB 31 & \quad  ABc 33  & \quad  ABc 31  \\
                                   &       &      &       &       &       &       &        &  \\ 
\noalign{\smallskip}\hline\noalign{\smallskip}

  0  \hspace{0.4cm}                &   12  &   25 &    7 &    9 \qq  &   7 &   5  & 2\qq   & 1 \qq \\
  5  \hspace{0.4cm}                &   12  &   25 &    7 &    9 \qq  &   7 &   5  & 3 ( 1)  & 1 \qq  \\
 10  \hspace{0.4cm}                &   14  &   24 &    8 &   10 \qq  &   8 &   6  & 3 ( 1)  & 1 \qq  \\
 15  \hspace{0.4cm}                &   17  &   24 &    9 &   11 \qq  &   9 &   6  & 3 ( 1)  & 1 \qq   \\
 20  \hspace{0.4cm}                &   22  &   23 &   10 &   12 \qq  &  11 &   7  & 4 ( 1)  & 1 \qq  \\
 25  \hspace{0.4cm}                &   32  &   21 &   12 &   15 \qq  &  16 &   9  & 5 ( 1)  & 1 \qq   \\
 30  \hspace{0.4cm}                &   55  &   17 &   15 &   19 \qq  &  26 &  11  & 7 ( 1) & 1 \qq   \\
 35   \hspace{0.4cm}               &  104  &   20 &   19 &   25 \qq  &  46 &  17  & 10 ( 1) & 2 \qq  \\
 40  \hspace{0.4cm}                &  203  &   43 &   25 &   33 \qq  &  88 &  26  & 14 ( 1) & 4 \qq  \\
 45 \hspace{0.4cm}                 &  394  &   97 &   36 &   47 ( 1) & 174 &  45  & 20 ( 1) & 6 ( 1) \\
 50  \hspace{0.4cm}                &  743  &  197 &   58 &   67 ( 2) & 345 &  85  & 32 ( 3) & 9 ( 2) \\
 55 \hspace{0.4cm}                 & 1307  &  362 &  101 &   86 ( 5) & 657 & 161  & 47 ( 2) & 15 ( 3)\\
 60 \hspace{0.4cm}                 &       &  593 &  180 &  105 ( 9) &1141 & 287  & 78 ( 2) & 27 ( 5)  \\
 65  \hspace{0.4cm}                &       &  862 &  309 &  113 (15) &     & 471  & 139 ( 2) & 52 ( 4)   \\
 70  \hspace{0.4cm}                &       & 1152 &  509 &  108 (24) &     & 725  & 255 ( 2) &  91 ( 4) \\
 75  \hspace{0.4cm}                &       &      &  822 &  102 (37) &     &1079  & 468 ( 4) & 163 ( 9)  \\
 80  \hspace{0.4cm}                &       &      & 1320 &  135 (33) &     &      & 856 ( 6) & 280 ( 9)   \\
 85  \hspace{0.4cm}                &       &      &      &  275 (37) &     &      & 1567 (10) & 511 (17)  \\
 90  \hspace{0.4cm}                &       &      &      &  488 (54) &     &      &          & 940 (44)   \\
\noalign{\smallskip}\hline

\end{tabular}
\ece

\end{table*}
%%%%%%%%%%%%%%%% end var results 
%\clearpage
\noindent
For the different
variational approximations we use the following naming scheme:
\be
\left. \begin{array}{rl}
\mbox{B 3d} \>     & : \> \mbox{variational {\it ansatz} with linear term} \> \>  \B \\
                   &   \> \> \>   \mbox{only \quad \cite{Carr}} \> , \\
\mbox{Bc 3d} \>    & : \> \mbox{variational {\it ansatz} with linear term} \> \>  \B \\   
                   &   \>  \> + \> \mbox{second cumulant \quad \cite{Carr}} \> ,\\
\mbox{AB 3d} \>    & : \> \mbox{variational {\it ansatz} with quadratic} \\ 
                   &    \>\>  \mbox{+ linear terms} \> \> \A \> \mbox{and} \> \B \> ,\\  
\mbox{ABc 3d} \>   & : \> \mbox{variational {\it ansatz} with quadr. + linear} \\
                   &   \>\> \> \mbox{terms} \>  \> \A \> \mbox{and} \> \B \> + \> \mbox{second cumulant} 
\> , 
\end{array} \right \}
\label{name var approx}\\
\ee
where $ d = 3$ (``aikonal'') or $d = 1$ (``ray''). Note that we list 
\be
10^4 \cdot \left | \frac{\Delta f}{f} \right | 
\ee
which is necessary to show the rather small deviations of the scattering amplitude -- 
any discrepancies in the cross sections 
wouldn't be seen on a logarithmic scale  except at large scattering angles.

Two independent programs using different subroutines and (partially) different techniques have been 
developed and run to arrive at these results. Many checks have been performed (see app. D) 
including a test of the Feyn\-man-Hellmann theorem for the interaction strength which is a highly 
nontrivial test of how well the variational equations are fulfilled.
In addition, the various accuracy parameters have been
varied to ensure stability of the numerical outcome. For the approximations including the second 
cumulant which demand particular care
we also include in table \ref{tab: var results} 
an approximate numerical error estimated from the different results from both programs and from variation of
the integration parameters.

%%%%%%%%%%%%%%%%%%%%% Fig. reldev_var
\begin{figure*}[hbt]
\vspace*{0.4cm}

\bce
\includegraphics[angle=90,scale=0.66]{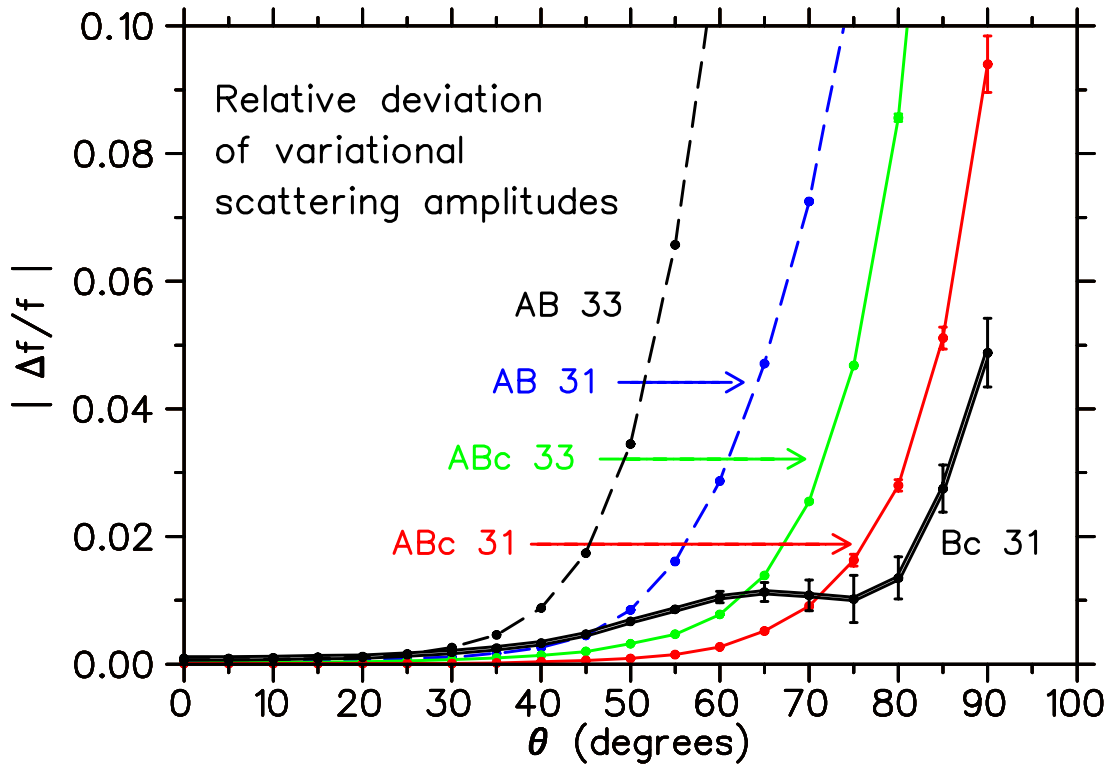}
\ece
\caption{(Color online) The relative deviation $|\Delta f/f|$ of the variational amplitudes from the
exact scattering amplitude as function of the scattering angle $\theta$. The naming scheme
(\ref{name var approx})
is used for the various approximations.}

\label{fig: reldev var}
\vspace*{1cm}
\end{figure*}
%%%%%%%%%%%%%%%%%%%%% end Fig. var reldev_var

%\clearpage

A systematic improvement is seen when quadratic terms 
and/or the correction by the second cumulant are included. 
This is shown in fig. \ref{fig: reldev var} were 
the deviations of the different results for the scattering amplitude are plotted
on  a {\it linear} scale. We also have included the best result from the variational approximation with 
a linear ansatz plus the second cumulant (Bc 31). 
It is seen that including the second cumulant gives much improved results so that the variational 
approximations labelled Bc 31 and ABc 31 give the best overall results.  However,
at present we do not fully understand why at larger scattering angles 
the (much simpler) scheme Bc 31 outperforms the more involved
scheme ABc 31 which includes the quantum-mechanical spreading effects. As we simply have added the 
second cumulant on top of the variational results and not used it in the  variational optimization
there is, of course, no guarantee that ABc 31 does {\it always} better than Bc 31 whereas AB 3d has always
to be better or at least as good as B 3d \footnote{Otherwise the variational principle would have chosen
the free $\A$ as the best solution.}. It should also be noted that the evaluation of the second cumulant
at large scattering angles where the amplitude is down by many orders of magnitude is a very demanding 
numerical task and the numerical errors given in table \ref{tab: var results} may be underestimated. 

Nevertheless, 
the full ``ray'' approximation plus the second cumulant (ABc 31), for example,  provides a 
very good approximation
even at larger scattering angles: at $\theta = 90^{\circ}$ the complex scattering amplitude is reproduced
within 9 \% and the cross section to better than 2 \% although the latter one is down by 5 orders 
of magnitude from the value taken in forward direction. 

Only at large scattering angles the deviations from the exact partial-wave result again start 
to grow and become so large that they even can be seen on the usual 
logarithmic plot for the cross section in fig. \ref{fig: cross-sec}.

\subsection{Comparison with other work}

Combining the eikonal approximation with the second-order Born approximation Chen has proposed 
several approximations which should also work at larger scattering angles and which we denote 
by ``Chen 1'' \cite{Chen1}, ``Chen 2'' \cite{Chen2} and ``Chen 3'' \cite{Chen3}. 

The relative deviation
of his approximate amplitudes from the exact value has been calculated by direct numerical integration 
and is displayed in table \ref{tab: Chen results} and fig. \ref{fig: reldev chen}. It is seen that 
these approximations are clearly inferior to the present variational approximations although they use
some analytic properties specific to the potential under consideration. In contrast, the variational approach
requires as input just the (Gaussian transform of the) potential and thus can be applied more generally.

We also include the results for Wallace's systematic 
eikonal expansion \footnote{For a Gaussian potential the 
explicit expressions are given in table I of ref. \cite{Wall}.} including real and imaginary phases 
up to second and third order in $1/k$. This makes for a relevant comparison with our variational results
which have been shown to be  correct up to second order in the high-energy expansion but contain 
many other higher-order terms. 

Table \ref{tab: Chen results} and fig. \ref{fig: reldev chen} show that
these systematic expansions are very good in forward direction but fail at higher scattering angles. Again
the variational approximations are clearly better in this kinematical regime.

\vspace{0.5cm}

%%%%%%%%%%%%%%%%%%%%% Fig. x-sec
\begin{figure*}[hbt]
\vspace*{0.5cm}

\bce
\includegraphics[angle=90,scale=0.65]{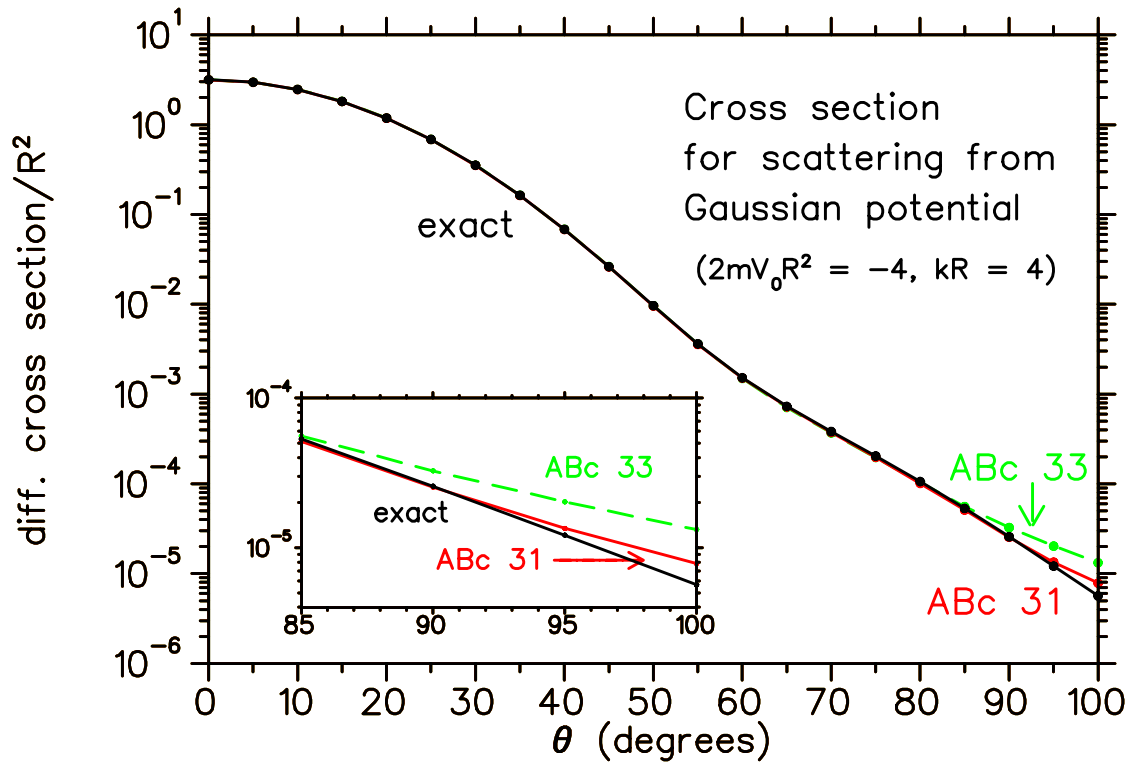}
\ece
\caption{(Color online) The exact cross section for scattering from a Gaussian potential with strength
$2m V_0 R^2 = -4$ at $ k R = 4 $ and the best variational cross sections including the second cumulant.}
\label{fig: cross-sec}
\vspace*{1cm}
\end{figure*}
%%%%%%%%%%%%%%%%%%%%% end Fig. x-sec

%%%%%%%%%%%%%%%%%  Table of Chen's results ( 7. 9. 09) and Wallace's expansion (9. 9.)

\begin{table*}[htb]
\caption{Relative deviation (in units of $10^{-4}$) of the scattering amplitude from the exact value
for various approximations by  Chen \cite{Chen1,Chen2,Chen3} and Wallace \cite{Wall}.
In the latter case the subscripts denote the order of the eikonal expansion which is included.
The potential is again a Gaussian one with 
the same parameters as in table \ref{tab: var results}.
}
\label{tab: Chen results}
\bce
\begin{tabular}{rrrrrr} \hline\noalign{\smallskip}
                     &          &          &             &             &          \\
\quad $\theta \quad  [ \, ^{\circ}\, ] $ & \quad Chen 1 &  \quad  Chen 2 & \quad  Chen 3   \quad  & 
\hspace*{0.6cm} Wallace$_2$    & \hspace*{0.6cm}  Wallace$_3$  \\
                     &          &          &             &             &          \\
\noalign{\smallskip}\hline\noalign{\smallskip}
  0  \hspace{0.4cm}  &    217   &      5   &   40 \quad  &    4 \quad  &  $<$ 1 \quad    \\
  5  \hspace{0.4cm}  &    220   &      6   &   40 \quad  &    4 \quad  &      1 \quad      \\       
 10  \hspace{0.4cm}  &    231   &      9   &   42 \quad  &    5 \quad  &      1 \quad      \\
 15  \hspace{0.4cm}  &    250   &     15   &   45 \quad  &    7 \quad  &      1 \quad      \\
 20  \hspace{0.4cm}  &    279   &     25   &   49 \quad  &    9 \quad  &      1 \quad       \\ 
 25  \hspace{0.4cm}  &    320   &     42   &   55 \quad  &    9 \quad  &      2 \quad      \\
 30  \hspace{0.4cm}  &    378   &     67   &   63 \quad  &    8 \quad  &      4 \quad       \\
 35  \hspace{0.4cm}  &    456   &    108   &   72 \quad  &    3 \quad  &      9 \quad       \\ 
 40  \hspace{0.4cm}  &    560   &    170   &   84 \quad  &   17 \quad  &     15 \quad       \\
 45 \hspace{0.4cm}   &    693   &    266   &   98 \quad  &   53 \quad  &     24 \quad     \\
 50  \hspace{0.4cm}  &    844   &    406   &  115 \quad  &  122 \quad  &     35 \quad      \\
 55 \hspace{0.4cm}   &    971   &    585   &  142 \quad  &  232 \quad  &     47 \quad      \\
 60 \hspace{0.4cm}   &   1001   &    767   &  194 \quad  &  376 \quad  &     66 \quad      \\ 
 65  \hspace{0.4cm}  &    891   &    907   &  281 \quad  &  528 \quad  &    110 \quad    \\
 70  \hspace{0.4cm}  &    672   &    999   &  405 \quad  &  678 \quad  &    192 \quad     \\
 75  \hspace{0.4cm}  &    400   &   1078   &  563 \quad  &  844 \quad  &    317 \quad      \\ 
 80  \hspace{0.4cm}  &    244   &   1190   &  765 \quad  & 1065 \quad  &    499 \quad      \\
 85  \hspace{0.4cm}  &    555   &   1390   & 1024 \quad  & 1392 \quad  &    763 \quad      \\
 90  \hspace{0.4cm}  &   1082   &   1739   & 1357 \quad  & 1898 \quad  &   1150 \quad      \\
% 95  \hspace{0.4cm}  &   1758   &   2301   & 1781 \quad  & 2676 \quad  &   1712 \quad      \\
%100  \hspace{0.4cm}  &   2590   &   3129   & 2309 \quad  & 3845 \quad  &   2517 \quad     \\
                     &          &          &             &             &            \\  
\noalign{\smallskip}\hline 
\end{tabular}
\ece
\vspace*{1cm}

\end{table*}
%%%%%%%%%%%%%%%% end Chen & Wallace results
%\clearpage

%%%%%%%%%%%%%%%%%% Fig. Chen_Wallace
\begin{figure*}[hbt]
\bce
\includegraphics[angle=90,scale=0.55]{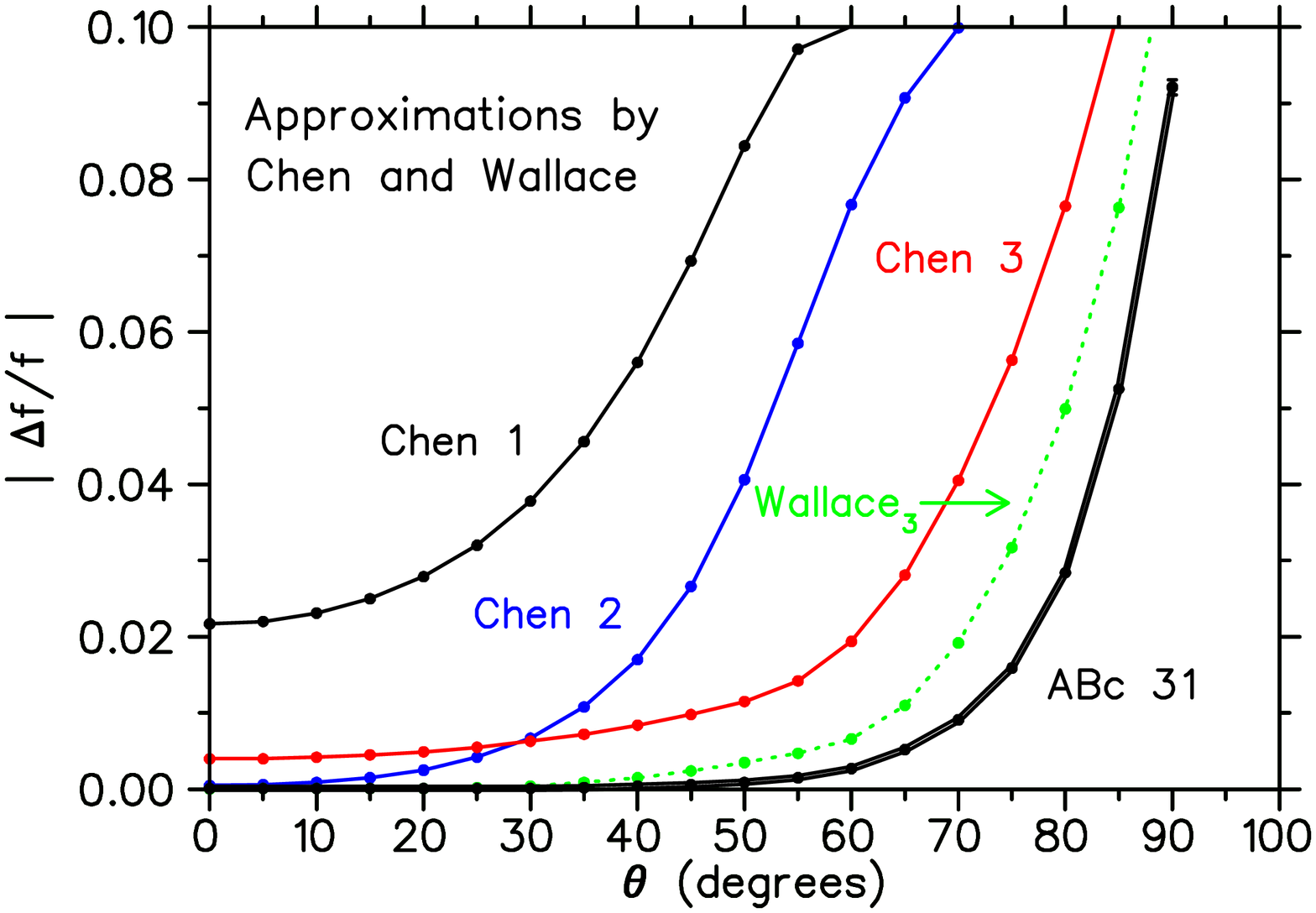}

\caption{(Color online) Same as in fig. \ref{fig: reldev var} but for the various approximations by 
Chen and Wallace's 3rd order eikonal expansion. For comparison also the best variational result 
(ABc~31 in the naming scheme (\ref{name var approx})) is shown.}
\label{fig: reldev chen}

\ece
\end{figure*}
%%%%%%%%%%%%%%%%%%%% end Fig. Chen_Wallace

%%%%%%%%%%%%%%%%%%%%%%%%%%%%%%%%%%%%%%%%%%%%%%%%%%%%%%%%%%%%%%%%%%%%%%%
%%% ************** SUMMARY ********************************************
%%%%%%%%%%%%%%%%%%%%%%%%%%%%%%%%%%%%%%%%%%%%%%%%%%%%%%%%%%%%%%%%%%%%%%%

\vspace{0.5cm} 

\section{Summary and outlook}

We have applied the Feynman-Jensen variational principle to two variants of a new path-integral 
representation of the ${\cal T}$-matrix in potential scattering 
using the most general {\it trial action} which contains quadratic and linear terms both in velocity 
and anti-velocity. Merging these variables into a ``super-velocity'' 
the variational functional was easily derived and the variational equations immediately followed
by demanding stationarity of that functional. As expected the anti-velocity (``phantom'') 
degrees of freedom prevented any divergence to appear when the scattering time was sent to infinity
as it should be when the initial and final states are detected in the scattering process.

We have found that this quadratic + linear {\it ansatz}
describes the scattering in terms of a (nearly) classical trajectory $\vecrho(t)$ with 
peculiar boundary conditions and a matrix operator $\Sigma$ which is the solution of a 
one-dimensional Lippmann-Schwinger-like equation. This reflects the classical as well as the 
quantum-mechanical (``wave spreading'') aspects of the process dominant in the high-energy,
forward scattering and low-energy scattering regime, respectively. 
% new 25. 8.
Unlike the previous linear {\it ansatz} \cite{Carr} the present variational approach
immediately leads to complex phases in the impact-parameter ${\cal S}$-matrix as required
by unitarity and a fluctuation
term coming from the corresponding functional determinant. We were also able to evaluate the
first correction (the second cumulant) in a systematic cumulant expansion.
%end new

As we solved the nonlinear, 
coupled variational equations
iteratively (both in analytic and numerical evaluations) the present work is mostly devoted
to the high-energy case with ``wave-spreading'' corrections becoming more important at 
backward scattering angles. Indeed we have seen better and better agreement with exact 
partial-wave calculations of scattering from a Gaussian potential as we step from the linear trial action
to the quadratic + linear and from the version which uses a straight-line reference path and 
a 3-dimensional anti-velocity (``aikonal'' or ``3-3'') to the version with a ray-like reference
path and a 1-dimensional anti-velocity (``ray'' or ``3-1''). An even more dramatic improvement 
is obtained if the correction by the second cumulant is added because then the second-order Born
approximation is contained in the variational result -- an observation also made by Chen 
(refs.  \cite{Chen1} - \cite{Chen3} ). 
% new 9.9.
Instead of plotting the approximate cross sections 
on a logarithmic scale as is usually done, 
we give the numerical values for 
the relative deviation of the approximate scattering amplitudes
from the exact partial-wave result.
This is done for a particular benchmark case: scattering from a Gaussian potential with 
fixed parameters (\ref{param}) and allows a {\it quantitative} comparison of the different approaches 
for a case where
large-angle scattering has been notoriously difficult to describe. 

We also have shown analytically that in a high-energy expansion terms up to inverse second order 
in $k$  or $K = k \cos(\theta/2)$ for ``ray'' or ``aikonal'' representation 
are correctly reproduced. Thus our variational approximation + cumulant correction is correct up to 
next-to-next-to-leading order in a systematic eikonal expansion.

However, there is no reason to limit the variational approach to high energies only since we expect
it to be useful at lower energies also. Indeed, it is just one of the inherent 
advantages of a variational description over an expansion in some small parameter
that the variational principle stretches the parameters/functions in the trial {\it ansatz}
in such a way as to make it ``optimal'' even under unfavorable kinematic conditions. Unfortunately,
we were not able to corroborate this expectation either by analytical or numerical methods. 
Obviously, a better strategy to solve the highly nonlinear variational equations is required.
\vspace*{0.2cm}

This is one of the tasks left for the future. The other one is, of course, to apply this
approximation scheme to more realistic situations, e.g. many-body scattering. Since the case of a 
spherically symmetric potential $V(r)$ was only chosen to compare with a (numerically) 
exactly solvable model, there is no difficulty to apply our variational scheme to general potentials
and scattering from few-body targets. A good step into that direction would be
to compare with the Faddeev results from ref. \cite{ELGJ} where $n+d$ scattering with a simple model 
potential has been evaluated
numerically for energies up to $1500$ MeV and compared with the Glauber approach. 
The path-integral representation and this variational approach will certainly be also useful 
in attempts to evaluate the scattering amplitude by stochastic (Monte-Carlo) methods \cite{Smir,MC-Ro}.
Finally, as the present approach relies solely on actions and not wave function(al)s
rendering it {\it a priori} more general, it seems that 
applications to nonperturbative field theory may also be possible.
\clearpage

\twocolumn[\hspace*{7.6cm}\huge\bf Appendix \\ \\]

%%%%%%%%%%%%%%%%%%%%%%%%%%%%%%%%%%%%%%%%%%%%%%%%%%%%%%%%%%%%%%%%%%%%%%%%%
%%% ******************** APPENDICES *************************************
%%%%%%%%%%%%%%%%%%%%%%%%%%%%%%%%%%%%%%%%%%%%%%%%%%%%%%%%%%%%%%%%%%%%%%%%%

\setcounter{section}{0}
\renewcommand{\thesection}{\Alph{section}}
\renewcommand{\thesubsection}{\thesection.\arabic{subsection}}
\renewcommand{\theequation}{\thesection.\arabic{equation}}
\newcommand{\adot}{\hspace*{-0.1cm}.\hspace*{0.4cm}}

%\vspace*{0.3cm}
%\bce
%{\huge\bf Appendix}
%\ece
%%%%%%%%%%%%%%%%%%% appendix A: averages
%%%%%%%%%%%%%%%%%%%
%\vspace*{0.2cm}

\section{\adot Evaluation of averages for the variational calculation}
\label{app: averages}
\setcounter{equation}{0}

In this section, we evaluate the different averages entering the Feynman-Jensen variational principle.
\vspace{0.2cm}

% new 16. 9.
We will be using a short-hand notation which simplifies the algebra considerably: first, as usual the 
convention  that over repeated indices is summed and second that continuous variables (times) are treated
the same way. More specifically, $i,j \ldots = 1,2,3 $ denote the cartesian components of vectors and 
matrices, $ \alpha, \beta \ldots = 1,2 \ldots (3+d) $ the components of super-vectors and 
$A \equiv (\alpha,t) $ etc. combine discrete and continuous indices. $\V^T, \Gamma^T \ldots $ denote the 
transpose of the corresponding quantity.
We do not distinguish between upper and lower indices.
Occasionally we also use the dot to denote a scalar product between vectors or between super-vectors. Thus, 
for example the trial action (\ref{A_t compact}) may be written in several different forms as
\bea
{\cal A}_t & \equiv & \frac{m}{2} \> \Iint dt' \> \sum_{\alpha,\beta=1}^{3+d} \V_{\alpha}(t) \, 
\A_{\alpha \beta}(t,t') \, V_{\beta}(t') \non
&& + \Iint \> \sum_{\alpha=1}^{3+d} B_{\alpha}(t) \, \V_{\alpha}(t) \non
\EA \frac{m}{2} \V_{\alpha}(t) \, \A_{\alpha \beta}(t,t') \, \V_{\beta}(t') +  
\B_{\alpha}(t) \, \V_{\alpha}(t) \non
\EA \frac{m}{2} \> \V_A \A_{AB} \V_B + \B_A \V_A 
\E   \frac{m}{2} \> \V^T \A \, \V + \B^T  \V \non 
\EA \frac{m}{2} \V \cdot \A \V + \B \cdot \V  
\eea
showing the obvious advantage of a concise notation.
% end new
\vspace{0.1cm} 

We first need the path integral over the trial action
\be
m_0 \> \equiv \> \int {\cal D}^3 v {\cal D}^d w \>  \exp \left ( \, i {\cal A}_t \, \right ) 
\ee
which from the rules of Gaussian integration can be immediately computed as
\be
m_0 \E \frac{\rm constant}{\mathrm{Det}^{1/2}(\A)} \, \exp\left(-\frac i{2m}\B\cdot\A^{-1}\B\right) \> .
\ee 
Here 
\be
\mathrm{Det} X \E \exp \left [ \, {\rm Tr} \, \ln X \, \right ] 
\ee
denotes a determinant, both in continuous (also called a functional determinant) 
and discrete variables and consequently ``Tr'' is a 
trace in all variables. The path integral in  eq. (\ref{PI for S(b)}) is normalized such that
it is unity without interaction, i.e. $\B = 0$ and $\A = \sigma_3 $. Here
$\sigma_3$ is the (extended) third Pauli matrix which appears
because the kinetic term of the anti-velocity has an opposite sign, 
i.e. $ \sigma_3 = {\rm diag}(1,1,1,-1,-1,-1)$
for the ``aikonal'' representation which uses a 3-dimensional anti-velocity and
%\vspace*{0.3cm}
%
$ \sigma_3 = {\rm diag}(1,1,1,-1) $ for the case of a 1-dimensional $w$.
This normalization determines the constant as $ {\rm Det}^{1/2} \sigma_3 $.
Defining for convenience the vector
\bea
&& \C \equiv \A^{-1}\B \> \> , \quad {\rm i.e.} \quad \> \> \C_A \E  \A^{-1}_{AB} \B_B \> \> , \non 
&& {\rm or} \quad \>  \> C_{\alpha}(t) \E
\Iint' \>  \A^{-1}_{\alpha \beta}(t,t')  \, \B_{\beta}(t')
\eea
and using the fact that $\sigma_3$ is its own inverse we finally have
\bea
m_0 \EA \mathrm{Det}^{-1/2}(\sigma_3\A) \, \exp\left(-\frac i{2m}\C\cdot\A\C \right) \non
\EA \exp \left [\,  - \frac{1}{2} {\rm Tr} \ln (\sigma_3 \A) -\frac i{2m}\C\cdot\A\C \, \right]
\> .
\label{result m_0}
\eea
This will also serve as a master integral or generating function for the averages we have to calculate.

The Feynman-Jensen variational principle requires 
the average of the difference between the full action and the trial action
\be
\la \Delta \cal A \rat  \E \frac m2\la \V\cdot(\sigma_3 - \A)\V\rat -\la\B\cdot \V\rat + 
\la\chi\rat \> .
\label{action diff}
\ee
Here
\be
\chi(\fb,\V] \E -\int_{-\infty}^{+\infty} dt \> V \left( \, \fx_{\mathrm{ref}}(t) + \fx_{\rm quant}(t) 
\, \right)
\label{phase chi}
\ee
is the phase in the path integral (\ref{PI for S(b)}) for the impact-parameter ${\cal S}$-matrix and 
the average has been defined in eq. (\ref{def average}).
%\vspace{0.5cm} 

\subsection{Computation of $\la \B\cdot \V\rat$}

The evaluation of this average is easily done by 
putting an artificial scalar factor $a$ in the linear term of the trial action and differentiating 
with respect to this factor
\bea
\la \B\cdot \V\rat \EA - i \frac d{da}  \ln m_0 (\C \to a \C)  \Biggr |_{a=1}  \non 
\EA - i \frac d{da}
\left(\,  -i\frac {a^2} {2m}\C\cdot\A\C \, \right)  \Biggr |_{a=1}  \non 
\EA-\frac 1 m \> \C \cdot \A \C \> .
\label{av linear}
\eea

\onecolumn
\subsection{Computation of $\frac m 2\la \V\cdot(\sigma_3 - \A)\V\rat$}

We obtain this average by the same technique: 
differentiating $m_0$ with respect to $\B$ brings down a factor $\V$. Therefore
\be
\frac m 2\la \V\cdot(\sigma_3 - \A)\V\rat \E -\frac 1{m_0}\frac m 2
\frac{\delta^2 m_0}{\delta\B_{\beta}(t')\delta\B_{\alpha}(t)} \, \Bigl [ \, \sigma_3^{\alpha \beta} 
\delta(t-t')
 - \A^{\alpha \beta}(t,t') \, \Bigr ] \> .
\ee
After the insertion of
\be
\frac{\delta^2 m_0}{\delta\B_{\beta}(t')\delta\B_{\alpha}(t)} \E m_0\left[-\frac 1{m^2} \C_{\alpha}(t)
\C_{\beta}(t') -\frac i m \A^{-1}_{\alpha \beta}(t,t')\right] 
\ee
we obtain
\be
\frac m 2\la \V\cdot(\sigma_3 - \A)\V\rat \E  -\frac m 2 \Biggl [-\frac 1{m^2} \C_A(t)\C_B(t') 
-\frac 1 m \A^{-1}_{AB}(t,t')\Biggr] \left[ \, \sigma_3^{AB}(t,t') - \A^{AB}(t,t') \, \right],
\ee
or, in continuous notation,
\be \label{ResE}
\frac m 2\la \V\cdot(\sigma_3 - \A)\V\rat \E \frac 1 {2m}\left[ \, \C\cdot\sigma_3\C - \C\cdot\A\C
\, \right] + \frac i2 \mathrm{Tr} \left[\, \sigma_3\A^{-1} - 1 \, \right] \> .
\ee
Together with eqs. (\ref{result m_0}) and (\ref{av linear}) we thus have for the variational functional
\be
S(\fb,\A,\C) \E \exp \Biggl \{ \, \frac{i}{2m} \C\cdot\sigma_3\C  
  - \frac{1}{2} \, {\rm Tr} \left [ \, \ln (\sigma_3 \A) + \sigma_3 \A^{-1} 
- 1  \, \right ] \> + i \, \la \chi \rat  \, \Biggr \} \> .
\label{S var 1}
\ee

%\vspace{0.5cm}

\subsection{Computation of $\la \chi \rat$}

Finally the average over the phase (\ref{phase chi}) can be done by a 
Fourier transformation of the potential: 
\be
\la \chi \rat  \E -\Iint \int \frac{d^3p}{(2\pi)^3} \> \widetilde V(\fp) \, \exp \left [ \, i \fp\cdot 
\fx_{\mathrm{ref}}(t) \, \right ] \> \la \exp \left ( \, i \fp \cdot \fx_{\rm quant}(t) \, \right ) \, 
\rat \> .
\label{<chi>1}
\ee
To simplify the calculation we write
\be
 \fx_{\rm quant}(t) \deF \Gamma(t,t') \, \V(t') \> \> , 
 {\rm i.e.} \quad x_{\rm quant}^i(t) \E
\Iint ' \> \Gamma_{i \alpha} (t,t') \V_{\alpha}(t')
\label{x from v}
\ee
where $\Gamma_{i \beta}(t,t')$ is a $ 3 \times (3+d)$-matrix operator. Note that as a messenger between
ordinary space and the ``super''-space in which velocity + anti-velocity live it cannot be a square matrix.
Indeed, from the connection (\ref{quant 33}) 
between quantum fluctuation and velocity we immediately find
% corrected terrible typo 18. 3. !
\be
\Gamma(t,t') \E \frac12\begin{pmatrix}\sgn(t-t')& 0 & 0 & -\sgn(-t')& 0&0 \\0& \sgn(t-t')&
 0 & 0 & -\sgn(-t')& 0 \\0&0&\sgn(t-t')& 0 & 0 & -\sgn(-t') \\ \end{pmatrix} \> ,
\label{Gamma 33}
\ee
for the ``aikonal'' case and from eq. (\ref{quant 31})
\be
\Gamma(t,t') \E \frac12 \begin{pmatrix}\sgn(t-t')-\sgn(-t')& 0 & 0&0\\0& \sgn(t-t')-\sgn(-t')& 0 & 0 \\0&
0&\sgn(t-t') & -\sgn(-t') \\ \end{pmatrix} \> ,
\label{Gamma 31}
\ee
for the ``ray'' case if we choose the mean momentum along the 3-axis.

In any case the average in eq. (\ref{<chi>1}) involves only  Gaussian integrals where in the exponent the
\be
\text{term linear in $\V$ } = \begin{cases} \quad i \, \left [ \, \B_{\alpha} + 
                                                  p_i \Gamma_{i \alpha} \, \right ]  
                                                 \V_{\alpha} & \text{for the numerator} \> ,\\
                          \quad i \, \B \cdot \V & \text{for the denominator.} \end{cases}
\ee
Thus
\bea 
\la \exp \left ( \, i \fp \cdot \fx_{\rm quant}(t) \, \right ) \, \rat \>  \EA \frac{1}{m_0} \, 
\int {\cal D}^3 {\cal D}^d w \> \exp \Biggl \{ \, i \, \frac{m}{2} \, \V^T \A V + 
i\left[ \, \B  + \fp \cdot \Gamma(t,\cdot) \, \right] \cdot \V \, \Biggr \} \non
\EA \exp \Biggl \{ \, -\frac i{2m} \left[ \, \B_{\alpha}(t') + p_i \, \Gamma_{i  \alpha}(t,t')
\,  \right ] \, \A^{-1}_{\alpha \beta}(t',t'') \, \left[\, \B_{\beta} (t'') + p_j \, 
\Gamma_{j \beta}(t,t'')\,  \right ] \, \Biggr \} \non
&& \times \exp \left  \{ \, \frac i{2m} \, \B \cdot \A^{-1} \B \, \right \} \> ,
\eea
where the last factor comes from the normalization of the average by $1/m_0$ and no summation/integration 
over the external time-parameter $t$ is implied.
Working out the exponential we find terms
linear and quadratic in $\fp$ while the constant term cancels.
The term linear in $\fp$ gives the trajectory
\be
\vecrho_i(t) \Def \fx_{\mathrm {ref}}^i(t) -\frac 1{m} \Gamma_{i \alpha}(t,t') \> 
\A^{-1}_{\alpha \beta}(t',t'') \> \B_{\beta}(t''),
\ee
or in short-hand notation
\be
\vecrho(t) \E \fx_{\mathrm {ref}}(t) -\frac1 m \Gamma(t,\cdot)\:\C .
\label{var rho 1}
\ee
The term quadratic in $\fp$ reads
\be
- \frac{i}{2m} \, p_i \,\Gamma_{i \alpha }(t,t') \> \A^{-1}_{\alpha \beta}(t',t'') \> 
\Gamma_{j \beta}(t,t'')  \, p_j 
\ee
which  becomes  $ - (1/2) p_i p_j \sigma_{ij}(t) $  if
a square $(3 \times 3)$-matrix
\be
\sigma_{ij}(t) \E  \frac i m\:\Gamma_{i \alpha}(t,t') \, \Gamma_{j \beta}(t,t'') \> 
\A^{-1}_{\alpha \beta }(t',t'') \> ,\quad\mathrm{or}\quad \sigma(t)
\E\frac i m\: \Gamma(t,\cdot)\A^{-1} \Gamma^T(\cdot,t)
\label{sigmat}
\ee
is defined (again no summation/integration over the external $t$).
From the definition (\ref{def Vsigma}) for the Gaussian transform of the potential we then obtain
the following simple expression for  $\la \chi \rat$ : 
\be 
\la \chi\rat \E -\Iint  \> V_{\sigma(t)}(\vecrho(t)).
\label{ResChi}
\ee

%%%%%%%%%%%%%%%%  appendix B : var. eqs.
%%%%%%%%%%%%%%%%
\section{\adot Variational equations}
\label{app: var eqs}
\setcounter{equation}{0}

In app. \ref{app: averages} we have calculated the quantities to be varied in the
Feynman-Jensen functional. 
The result (\ref{S var 1}) may be written as
\be
S(\fb,\A,\C) \E \exp\left[ \, i X_0 + i X_1 -\Omega(\sigma_3\A) \, \right],
\ee
with
\bea
X_0 &\equiv& \la \chi \rat \E  -\Iint \> V_{\sigma(t)}(\vecrho(t)), 
\label{def X0}\\
X_1 &:=&  \frac{1}{2m}\C\cdot \sigma_3\C,
\label{def X1}\\
\Omega(X) &:=&  \frac 12 \mathrm{Tr} \left[ \, \ln X + X^{-1} -1 \, \right] \> .
\label{def Omega}
\eea
The argument of the potential -- the trajectory -- is given in eq. (\ref{var rho 1}) and its Gaussian 
width in eq. (\ref{sigmat}). 
We will now compute the variational equations for $\A$ and $\B$ which follow when
stationarity of $S(\fb,\A,\B)$ is required.
However, since only the combination $\C = \A^{-1}\B \> $ (but not $\B$) and $\A^{-1}$ enter the 
functional to be varied, it is more convenient to vary with respect to $\C$ and $\A^{-1}$.

\subsection{Variational equation for $\C$}

The vector $\C$ only shows up in the phase $X_1$ and the trajectory $\vecrho(t)$. Therefore 
stationarity of the variational functional requires
\be
\frac{\delta X_1}{\delta\C_{\alpha}(t)} - \Iint' \> \partial_i V_{\sigma(t')} (\vecrho(t')) \> 
\frac{\delta \vecrho_i(t')}{\delta\C_{\alpha}(t)} \> \stackrel{!}{=} \> 0 
\ee
where we have used the chain rule in the last term. 
From eq. (\ref{def X1}) we immediately have
\be
\frac{\delta X_1}{\delta\C_{\alpha}(t)} \E \frac 1 m \, \left ( \sigma_3 \C \right)_{\alpha} (t)
\ee
and from eq. (\ref{var rho 1})
\be
\frac{\delta \vecrho_i(t')}{\delta\C_{\alpha}(t)} \E -\frac 1 m \Gamma_{i \alpha}(t',t) \> .
\ee
Thus
\be
\left ( \sigma_3 \C \right)_{\alpha}(t) \E - 
\partial_i V_{\sigma}(t') \, \Gamma_{i \alpha}(t',t) \> .
\ee
As $\sigma_3$ is its own inverse and diagonal in the time variables, we may
solve this as
\be
\C_{\alpha}(t) \E -  \left ( \sigma_3 \right )_{\alpha \beta} \, \partial_i V_\sigma(t')
\, \Gamma_{i \beta }(t',t),
\ee
or
\be
\C_{\rm var} \E - \sigma_3 \, \Gamma^T \, {\bf J}_{\sigma}
\label{C var}
\ee
where 
\be
{\bf J}_{\sigma}(t) \E  \nabla V_{\sigma(t)}(\vecrho(t))
\ee
is the vector of potential derivatives (also called the Jacobian).
We can now put this expression into the trajectory to get
\bea
\vecrho_{\rm var}(t) \EA \fx_{\mathrm{ref}}(t) - \frac 1 m \Gamma(t,s) \, \C^{\rm var} (s) \E 
  \fx_{\mathrm{ref}}(t) +\frac{1}{m} \Gamma(t,s) \sigma_3 \, \Gamma^T (s,t') \, {\bf J}_{\sigma}(t') \\
&=:& \fx_{\mathrm{ref}}(t) + \Sigma_0(t,t') \,  {\bf J}_{\sigma}(t')  \> .
\label{eq motion}
\eea
Here we have defined the $(3 \times 3)$-matrix
\be
\Sigma_0(t,t') \Def \frac{1}{m} \Gamma(t,s) \sigma_3 \, \Gamma^T (s,t')
\label{def Sigma0}
\ee
whose explicit form can be worked out with the help of eqs. (\ref{Gamma 33}) and (\ref{Gamma 31}) and 
the relation
\be
\int_{-T}^{+T} ds \>  \, \sgn(s-t) \, \sgn(s-t') \E 2 \left [ \>  T - |t-t'| \> \right ] \> .
\ee
Due to the contribution from the anti-velocity all terms which would diverge for $T \to \infty$ cancel 
and one obtains the expressions given in eqs. (\ref{Sigma0 33}) and (\ref{Sigma0 31}).
Note that $\Sigma_0$  is the Green function for the corresponding equation of motion. For example, we have
\be
 - m \, \frac{\partial^2}{\partial t^2} \,
\left ( \, \Sigma_0^{(3-3)} \, \right )_{ij} (t,t') = \delta(t-t') \, \delta_{ij}
\ee
so that one can write 
\be
\Sigma_0^{(3-3)} \E {\cal P} \, \frac{1}{-m \partial^2_t}
\ee
where ${\cal P}$ denotes the principal value prescription. Indeed,
evaluating in Fourier (E-) space with
$ (E|t) = \exp( i E t)/\sqrt{2 \pi} $ as transformation function we obtain
\bea
\left ( \, \Sigma_0^{(3-3)} \, \right )_{ij}(t,t') \EA \frac{\delta_{ij}}{m} \, 
\frac{1}{2 \pi} \, {\cal P} \,
\int_{-\infty}^{+\infty} dE \>
\frac{\exp ( i E (t-t'))}{E^2} \> \equiv \>   \frac{\delta_{ij}}{m} \, \frac{1}{2 \pi} \, 
\int_{-\infty}^{+\infty} dE \> \frac{\exp ( i E (t-t')) - 1}{E^2} \non
\EA -\frac{\delta_{ij}}{m} \, \frac{1}{\pi}  \int_0^{\infty} dE \>
\frac{1 - \cos(E(t-t'))}{E^2} \E -\frac{1}{2m} \, |t - t'| \, \delta_{ij} \> ,
\eea
in agreement with eq. (\ref{Sigma0 33}).

% corrected 14. 9.
\subsection{Variational equation for $\A^{-1}$}
\newcommand{\dec}{\hspace{0.2cm}}
The matrix $\A^{-1}$ only enters in the Gaussian width $\sigma$ (appearing in $X_0$)
and in the ``fluctuation term'' $\Omega$. Therefore requiring stationarity means
\be 
- \frac{\delta\Omega}{\delta \A^{-1}_{\alpha \beta}(t_1,t_2)} - i \Iint \> 
\frac{\delta V_{\sigma}(t)}{\delta \A^{-1}_{\alpha \beta}(t_1,t_2)} \> \stackrel{!}{=} \> 0 \> .
\label{var A 1}
\ee
From the definition  (\ref{def Vsigma}) we find
\be
\frac{\delta V_{\sigma}(t)}{\delta \A^{-1}_{\alpha \beta}(t_1,t_2)} \E 
\frac{1}{2} \, \partial_i \partial_j V_{\sigma}(t) \, 
\frac{\delta \sigma_{ij}(t)}{\delta \A^{-1}_{\alpha \beta}(t_1,t_2)} \> .
\label{Vsigma diff}
\ee
Thus, we have to compute the functional derivative of the Gaussian width and fluctuation term
with respect to  $\A^{-1}$.
This is easily done for the Gaussian width: From eq. (\ref{sigmat}) it follows:
\be
\frac{\delta\sigma_{ij}(t)}{\delta\A^{-1}_{\alpha \beta}(t_1,t_2)} \E \frac i m 
\Gamma_{i \alpha}(t,t_1)\Gamma_{j \beta}(t,t_2) \>.
\label{sigma diff}
\ee
For the fluctuation term we rewrite  eq. \eqref{def Omega} as
\be
\Omega(\sigma_3\A) \E
\frac 12 \mathrm{Tr}\left[ \, -\ln (\A^{-1}\sigma_3) + (\A^{-1}\sigma_3) - 1 \, \right]
\ee
and use the chain rule
(in our condensed notation from app. \ref{app: averages}) 
\be
\frac{\partial\Omega(\A^{-1} \sigma_3)}{\partial\A^{-1}_{AB}} \E 
\frac{\partial\Omega(\A^{-1} \sigma_3)}{\partial[\A^{-1}\sigma_3]_{CD}} \> 
\frac{[\partial \A^{-1}\sigma_3]_{CD}}{\partial\A^{-1}_{AB}} \E 
\frac{\partial\Omega(\A^{-1} \sigma_3)}{\partial[\A^{-1}\sigma_3]_{CD}} \> \delta_{AC} 
\left ( \sigma_3 \right )_{BD}\> .
\ee
Since \footnote{This can be easily proved by a power series expansion of 
$f(X)$, piecewise differentiation and resumming.} 
\be 
\frac{\partial \, \mathrm{Tr} f(X)}{\partial X_{CD} } \E \left [ f'(X) \right]_{DC}
\ee
for any differentiable function $f(X)$ we obtain
\be
\frac{\partial \Omega(\A^{-1}\sigma_3)}{\partial \A^{-1}_{AB}} \E
 -\frac 12 \left\{ \, \left ( \left [ \A^{-1} \sigma_3 \right ]^{-1} \right )_{DC} - \delta_{AD} \, 
\right\} \,\delta_{AC} 
\left ( \sigma_3 \right )_{BD} \E -\frac 12 \left\{ \, \A_{AB} - [\sigma_3]_{AB} \, \right\},
\ee
where we used the fact both $\sigma_3$ and $\A$ are symmetric.
In extended notation the results reads
\be
\frac{\delta\Omega(\sigma_3\A)}{\delta\A^{-1}_{\alpha \beta}(t_1,t_2)} 
\E -\frac 12 \left\{ \, \A_{\alpha \beta}(t_1,t_2) -  \left ( \sigma_3 \right)_{\alpha \beta} \, 
\delta \left ( t_1 - t_2 \right ) \, \right \} \> .
\ee
Utilizing eqs. (\ref{Vsigma diff}), (\ref{sigma diff}) the variational equation (\ref{var A 1}) 
therefore becomes
\be
-\frac 12 \left\{\A_{\alpha \beta }(t_1,t_2)- \left ( \sigma_3 \right )_{\alpha \beta} \, 
\delta \left ( t_1 - t_2 \right ) \, \right \} 
\E \frac 1 {2m} \,  \Iint \> \partial_i \partial_j V_{\sigma}(t) \Gamma_{i \alpha}(t,t_1)
\Gamma_{j \beta}(t,t_2).
\ee
In matrix notation this simply reads
\be
\A_{\rm var} \E \sigma_3 - \frac 1{m} \, \Gamma^T \, H_{\sigma} \, \Gamma 
\label{var A 2}
\ee
where $H_\sigma$ stands for the matrix of second derivatives (also called the Hessian) of the potential 
\be
H^{ij}_{\sigma} (t,t') \Def \partial_i \partial_j \, V_{\sigma(t)}(\vecrho(t)) 
 \> \delta \left ( t - t' \right ) \> .
\ee
It is possible to put the results in an even more compact form by defining
\be
\Sigma \Def \frac 1 m \Gamma \A^{-1} \Gamma^T \> .
\label{def Sigma}
\ee
Note that for the free case $\A = \sigma_3$ this reduces to the matrix $\Sigma_0$ defined in eq.
(\ref{def Sigma0}) which is the Green function for the equation of motion (\ref{eq motion}).
% style correction 14. 9.
Multiplying eq. (\ref{var A 2}) from left by $\Gamma \sigma_3$ and from right by $ \A^{-1} \Gamma^T/m $, 
the variational equation for $\A^{-1}$ then translates into a Lippmann-Schwinger-like integral equation
\be
\Sigma \E \Sigma_0  + \Sigma_0 H_{\sigma} \Sigma \> .
\ee

All relevant quantities can be written in terms of $\Sigma(t,t')$ and the variational trajectory 
$\vecrho(t)$. For instance from eqs. (\ref{def X1}), (\ref{C var}) and (\ref{def Sigma0}) we immediately 
obtain
\be
X_1 \E \frac 1 2 \, {\bf J}_{\sigma} \, \cdot \, \Sigma_0 \, {\bf J}_{\sigma} \> .
\ee
Less obvious is that the  fluctuation term  $ \Omega(\sigma_3 \A)$ defined in eq. (\ref{def Omega}) 
may also be expressed in terms
of $\Sigma$ and $\vecrho$. However, if the variational solution (\ref{var A 2}) is multiplied by $\A^{-1}$
and inserted into
\be
\Omega(\sigma_3 \A) \E - \frac{1}{2} {\rm Tr} \, \left [\, \ln \left (\A^{-1} \sigma_3 \right ) - 
\A^{-1} \sigma_3 + 1  \, \right ]
\ee
one obtains
\be
\Omega_{\rm var} \E - \frac{1}{2} {\rm Tr} \, \left [\, \ln \left ( 1 + 
\A^{-1} \frac{1}{m} \Gamma^T H_{\sigma} \Gamma \right ) - 
\frac{1}{m} \A^{-1} \Gamma^T H_{\sigma} \Gamma \, \right ] \> .
\ee
Using the definition (\ref{def Sigma}) and the properties of the trace this becomes
\be
\Omega \E - \frac 12 \mathrm{Tr} \left[ \> \ln (1 + \Sigma H_\sigma) - \Sigma H_\sigma \, \right] \> .
\ee
Finally from eq. (\ref{sigmat}) we see that the Gaussian width $\sigma(t) $ 
is just given by the diagonal element of $ i \Sigma $.

%%%%%%%%%%%%%%%%%%%%%%%%% appendix C: second cumulant
%%%%%%%%%%%%%%%%%%%%%%%%%
% changed 14. 9.
\section{\adot Calculation of the second cumulant}
\label{app: cum2}
\setcounter{equation}{0}
In this section we compute the first correction to the variational approximation, i.e. the second cumulant
\bea
\lambda_2 \EA \la \left(\Delta \cal A\right)^2 \rat - \la \Delta \cal A \rat^2 \\ 
\Delta{\cal A} \EA \frac m 2 \V\cdot\left(\sigma_3 -\A\right)\V - \B\cdot \V + \chi \> ,
\eea
for both representations. Since the computational steps are similar to the ones in app.
\ref{app: averages} we will not enter into too much details.
After some algebra we find that the second cumulant is given by the following terms which we will 
evaluate in the subsequently:
\bea
\lambda_2 \EA \la \chi^2 \rat - \la \chi \rat^2 
+ \frac{i}{m} \C \cdot \A\C + i m \left. \frac{d}{da} \right|_{a=1}
\left[ \, \la E_a \rat + \frac 2 m \la\chi_a \rat \, \right] 
- \left[ \, \left (\sigma_3 \right )_{\alpha\beta}(t,t') - \A_{\alpha \beta}(t,t') \, \right] \non
&& \quad \times \left[ \, 2\frac m {m_0}\frac{\delta m_0}{\delta\B_{\alpha}(t)}
\frac{\delta \la\chi\rat}{\delta\B_{\beta}(t')}  
+ m \frac{\delta^2 \la\chi\rat}{\delta\B_{\alpha}(t) \delta\B_{\beta}(t')}
+\frac {m^2} {2m_0}\frac{\delta m_0}{\delta\B_{\alpha(t)}} \frac{\delta \la E \rat}{\delta\B_{\beta}(t')} 
+  \frac {m^2}{4}\frac{\delta^2 \la E \rat}{\delta\B_{\alpha}(t) \delta\B_{\beta}(t')}
\, \right],
\eea
where the quantity $E$ stands for
\be
\la E \rat \Def \la \V\cdot \left(\sigma_3 -\A\right) \V \rat \> .
\ee
From our study of the first cumulant, eqs. (\ref{ResE}) and (\ref{ResChi}), we can compute
\be
\left.\frac d{da}\right|_{a = 1} \la \chi_a\rat \E \frac 1 m \nabla V_\sigma\cdot\Gamma \C,
\ee
and
\be
\left.\frac d{da}\right|_{a = 1} \la E_a\rat \E \frac 2{m^2}\left[\C\cdot\sigma_3\C -\C\cdot \A\C\right].
\ee
Similarly, we have
\bea
\frac 1 {m_0}\frac {\delta m_0}{\delta \B_{\alpha}(t)} \EA -\frac i m \C_{\alpha}(t)\\
\frac {\delta \la \chi\rat}{\delta \B_{\alpha}(t)} \EA \frac 1 m \left[ \, \nabla V_\sigma\cdot
\Gamma\A^{-1}\, \right]_{\alpha}(t) \\
\frac {\delta \la E\rat}{\delta \B_{\alpha}(t)}\EA \frac 2 {m^2} \left[ \, 
\C\cdot\sigma_3\A^{-1} -\C \, \right]_{\alpha}(t),
\eea
while
\bea
\frac{\delta^2 \la E \rat}{\delta\B_{\alpha}(t)\delta \B_{\beta}(t')} \EA \frac 2 {m^2}\left[ \, 
\A^{-1}\sigma_3\A^{-1} -\A^{-1} \, \right]_{\alpha \beta}(t,t') \\
\frac{\delta^2 \la \chi \rat}{\delta\B_{\alpha}(t)\delta\B_{\beta}(t')} \EA -\frac 1 {m^2}
\left[ \, \A^{-1}\Gamma^T H_\sigma
\Gamma\A^{-1} \, \right]_{\alpha \beta}(t,t') \\
&\stackrel{\mathrm{var}}{=}& -\frac 1 {m} \left[ \, \A^{-1}\left(\sigma_3 -\A\right)\A^{-1} \, 
\right]_{\alpha \beta}(t,t').
\eea
Here, the ``var'' above the equality sign in the last line means that the variational 
solution (\ref{var A 2}) has been used.
We have now everything needed to compute the second cumulant.
After some (tedious) algebra we obtain
\be
\lambda_2 \E \la\chi^2\rat - \la\chi\rat^2 -\frac i m \C\cdot \sigma_3\A^{-1}\sigma_3\C + 
\frac 12 \mathrm{
Tr} \left[\A^{-1}\sigma_3\A^{-1}\sigma_3 -1\right]- \mathrm{Tr} \left[\A^{-1}\sigma_3 -1\right].
\ee
% correct typo 14. 9.
Using the variational equations and the Lippmann-Schwinger-like equation we find
\bea
\mathrm{Tr}\left[\sigma_3\A^{-1} -1\right] \EA  \mathrm {Tr}\left[ \, \Sigma H_\sigma \, \right] \\
\frac 12\mathrm{Tr}\left[\, \sigma_3\A^{-1}\sigma_3\A^{-1} -1 \, \right] \EA  
\mathrm {Tr}\left[ \, \Sigma H_\sigma +
\frac 12  \left(\Sigma H_\sigma\right)^2 \, \right],
\eea
while
\be
\frac i m \C\cdot \sigma_3\A^{-1}\sigma_3\C \E i\nabla V_\sigma^T \, \Sigma \, \nabla V_\sigma,
\ee
so that our final result is
\be
\lambda_2 \E \la\chi^2\rat - \la\chi\rat^2 -i \nabla V_\sigma^T \, \Sigma \, \nabla V_\sigma
+ \frac 12 \mathrm{Tr} \left[ \Sigma H_\sigma \right]^2.
\ee
% style correction 14. 9.
The quantity $\la\chi^2\rat$ is evaluated in the very same way as $\la\chi\rat$, i.e. one 
transforms the potential to Fourier space, and performs the functional integrations. The result is
% shortened 14. 9.
the one given in eq. (\ref{<chi2>}).

%\vspace{1cm}
%%%%%%%%%%%%%%%%%%%%%%%%%%%%%%% appendix D: numerical details
%%%%%%%%%%%%%%%%%%%%%%%%%%%%%%%

\section{\adot Some numerical details}
\label{app: num details}
\setcounter{equation}{0}

Here we discuss several points which are essential to obtain reliable numerical values for our
variational approximations.

\subsection{Variables}

As the range of the potential greatly determines the dynamics of the scattering process
we switch from time variables to distances by substituting
\be
t \E \frac{z}{v_{\rm char}}
\ee
where the characteristic velocity is chosen as the asymptotic velocity on the different
reference trajectories, {\it viz.}
\be
v_{\rm char} \E \left \{ \begin{array}{r@{\quad:\quad}l}
                         \frac{K}{m}: & {\rm ``aikonal''} \\
                         \frac{k}{m}: & {\rm ``ray''} \> .\end{array} \right .
\ee
This is reasonable for high-energy scattering where indeed the particle mostly travels along
the reference trajectory and implies that each power of $t$ and each integration over $t$
is suppressed by an inverse power of $K$ (or $k$). It becomes less convincing at low energy
where the characteristic velocity may be totally different from the asymptotic one - for example,
in scattering via a resonance. However, all our applications will be in the high-energy domain.

\twocolumn
\subsection{Numerical integration}

Initially our numerical integrations have been performed 
by using Gauss-Legendre quadrature rules 
for a {\it finite} interval $ \> [- z_{\rm max},+z_{\rm max}] \> $. Typically we have taken 
\be
z_{\rm max} \> \sim \>  ( 4 \div 6 ) \cdot R
\ee
as we expect that outside this interval the potential is practically zero. Of course,  
$z_{\rm max}$ had to be varied to ensure stable results. In previous work we had 
mapped the {\it infinite}
$z$-interval to a finite one (for example by $ z = R \tan \psi$) but in the present case 
this caused some problems for the iterative solution of the variational equations 
% new 6. 10. 
which are avoided if the integration is over a finite, not too large interval.	

The numerical integration was performed with $n_e$ subdivisions and
$n_g$ Gaussian points in each subinterval 
% new 6. 10.
requiring $ N = n_e \times n_g $ function calls altogether.
Again the number $n_e$ was varied to verify stability of the
results whereas the number $n_g$ was kept fixed at moderate values which is 
advisable for oscillatory integrands, e.g. $(n_g,n_e) = (32,2)$. Since we deal with 
multi-dimensional integrals and have to solve the variational equations for each $\fb$-value 
on the grid the computing time rapidly increases when making the grid finer and finer and 
at large scattering angles (where the scattering amplitude becomes small by interference) 
it became difficult to obtain stable numerical results. In these cases use of the 
adaptive routine DCUHRE \cite{Cuhre,Hahn} 
%new 14. 12. 
for the $\fb$-integration was quite helpful as it chooses the integration points
according to their relative importance. 

% new 5. 10. 
\vspace{0.1cm}

Later we realized that integrating numerically over nonanalytic functions 
like $ |z-z'| $ or $ |z'| $ 
(which occur in the Green function $ \Sigma_0(z,z') $ ) is not well handled by Gaussian or similar
quadrature rules. The reason is that their error is proportional to some high
derivative of the integrand within the integration interval which makes them suitable for 
analytic functions 
but not for nonanalytic ones. Consequently, the simple trapezoidal rule (see, e.g. eq. (25.4.2) in 
ref. \cite{Handbook}, with $h = (z_N - z_0)/N $ denoting the increment)
\bea
\int_{z_0}^{z_N} dz \> f(z) \EA h  \, \left [ \, \frac{f_0}{2} + f_1 + \ldots + f_{N-1} + 
\frac{f_N}{2} \, \right ] \non
&& - \frac{N \,  h^3}{12} \, f''(\xi) \> , \quad z_0 < \xi < z_N
\eea
is as efficient (or better: inefficient) in integrating nonanalytic 
functions as a $N$-point Gaussian integration which is exact for polynomials up to $z^{2N-1}$ or 
Simpson's rule whose error is proportional to $h^5 f^{(4)}(\xi)$. This is demonstrated in table 
\ref{tab: int_abs} where the test cases
\bea
\hspace*{-0.5cm} I_1(z) \! &:=& \! \int\limits_{-\infty}^{+\infty} dz'  \, | z - z' |  e^{-z'^2} 
= \sqrt{\pi}  z  \cdot {\rm erf}(z) + e^{-z^2} 
\label{I1} \\
\hspace*{-0.5cm} I_2(x) &:=&  \int\limits_{-\infty}^{+\infty} dz' \> \Bigl \{ \, | z - z' | - | z | - 
| z' | \, \Bigr \} \, 
e^{-z'^2} \non
\EA I_1(z) - \sqrt{\pi} \, | z | - 1
\label{I2}
\eea
are considered as typical examples of integrals for a Gaussian potential in the ``aikonal'' and 
the ``ray'' representation, respectively. Here erf$(z)$ denotes the standard error function with
${\rm erf}(-z) = - {\rm erf}(z) $.
It is obvious that the accuracy is relatively poor 
and the convergence with increasing number $N$ of function calls is disappointingly slow~\footnote{For 
comparison: without absolute sign in the integrand, i.e. for an analytic function, 
the absolute deviation from the exact result is lower than $10^{-11}$ already 
at  $ N = 48 $ for all integration rules. Strictly speaking, of course,   
the equally spaced rules require $N+1$ function calls.}. This can be improved in the following way:
\vspace{0.2cm}

%%%%%%%%%%%%%%%%%  Table int_abs

\begin{table*}[htb]
\caption{Maximal deviation of the numerically evaluated integrals $I_{1/2}(z)$ 
(see eqs. (\ref{I1}), (\ref{I2})) from the exact value   
 for different quadrature rules and function calls $N$ 
in the interval $ z \in [-5,5]$. The integration range was made finite by
cutting off the integrand for $|z'| > 5$. ``Gauss 24'' etc. refers to a Gauss-Legendre quadrature 
rule with $n_g = 24$ points etc. The last two lines give the
results obtained with the Euler-MacLaurin summation formula retaining corrections up to second and
fourth power in the increment, respectively, due to the nonanalytic behaviour of the integrand.
} 
\label{tab: int_abs}
\bce
\begin{tabular}{lcccc|cccc} 
\hline\noalign{\smallskip}
                                 &      &      &      &       &      &      &      & \\

                                 & \multicolumn{4}{c|}{$I_1(z)$} & \multicolumn{4}{c}{$I_2(z)$} \\
                                 &      &      &      &       &      &      &      & \\
\quad Rule \quad N $\rightarrow$ & $24$ & $48$ & $72$ & $144$ & $24$ & $48$ & $72$ & $144$ \\
\quad \quad $\downarrow$         &      &      &      &       &      &      &      & \\
%                                 &      &      &      &       &      &      &      & \\ 

\noalign{\smallskip}\hline\noalign{\smallskip}
                                &      &      &      &       &      &      &      & \\ 

Gauss 24 &  6.38 (-2) & 5.12 (-3) & 7.53 (-3) & 1.24 (-3) & 1.01 (-1) & 5.12 (-3) & 1.14 (-2) & 1.24 (-3) \\

Gauss 48 &            & 1.72 (-2) &           & 1.94 (-3) &           & 2.61 (-2) &           & 2.91 (-3) \\

Gauss 72 &            &           & 7.77 (-3) & 5.80 (-4) &           &           & 1.17 (-2) &  5.78 (-4) \\

Simpson  &  5.04 (-2) & 1.40 (-2) & 6.34 (-3) & 1.60 (-3) & 5.30 (-2) & 1.42 (-2) & 6.37 (-3) &  1.60 (-3) \\

Trapez   &  2.95 (-2) & 7.27 (-3) & 3.22 (-3) & 8.04 (-4) & -2.95 (-2) & -7.27 (-3) & -3.22 (-3) & -8.04 
(-4) \\ 
                                 &      &      &      &       &      &      &      & \\ \hline
                                 &      &      &      &       &      &      &      & \\ 

Euler-MacLaurin$_2$ & 5.25 (-4) & 3.17 (-5) & 6.23 (-6) & 3.88 (-7) & -7.57 (-4) & -4.59 (-5) & -9.01 (-6) & 
-5.61 (-7) \\

Euler-MacLaurin$_4$ & 6.33 (-5) & 1.00 (-6) & 8.81 (-8) & 1.39 (-9) & -9.91 (-5) & -1.60 (-6) & -1.43 (-7) & 
-2.23 (-9) \\ 
                                 &      &      &      &       &      &      &      & \\ 
\noalign{\smallskip}\hline

\end{tabular}
\ece
\vspace*{0.8cm}
\end{table*}
%%%%%%%%%%%%%%%%%%%%  end table int_abs

\noindent
For simplicity, we first consider integrands of the $I_1$-type, i.e.
\be
f(z') \E |z - z'| \, \phi(z') + g(z')
\label{full integrand}
\ee
where $\phi(z'), g(z')$ are differentiable functions which vanish rapidly enough for $ z' \to \pm \infty$. 
Therefore to a good approximation
\bea 
&& \int\limits_{-\infty}^{+\infty} dz' \, \Bigl \{ \, | z - z' | \, \phi(z') + g(z') \, \Bigr \}   \simeq  
\int\limits_{-z_{\rm max}}^z dz' \,  \Bigl [ \, \left ( z - z' \right ) \, \phi(z') \non 
&& \hspace*{1cm} + g(z') \, \Bigr ] 
+ \int\limits_z^{+z_{\rm max}} dz' \,  \Bigl [ \, \left ( z' - z \right ) \, \phi(z') 
+ g(z') \, \Bigr ] \> .
\eea
Note that the integrand is analytic but different in the two integrals.
We may now apply the Euler-MacLaurin summation formula (ref. \cite{Handbook}, eq. (25. 4. 7))
assuming that
the point  of nonanalyticity $ z = k \, h $ is a multiple of the increment $ h = 2 z_{\rm max}/N $. As we 
are evaluating the variational equations iteratively on a grid this certainly is the case and we obtain
\bea
&&\int\limits_{-z_{\rm max}}^{+z_{\rm max}} dz' \> \Bigl \{  \, | z - z' | \, \phi(z') + g(z') \, \Bigr \} 
\, \simeq \,  \mbox{ trapezoidal rule} \non
&& \quad- \sum_{m=1}^n \, \frac{B_{2m} h^{2m}}{(2m)!} \, \Biggl \{  \, 
\frac{d^{2m-1}}{dz'^{2m-1}} \Bigl [ \, (z-z') \phi(z') + g(z') \, \Bigr ] \non
&&  \hspace*{2cm} - \, \frac{d^{2m-1}}{dz'^{2m-1}}
\Bigl [ \, (z'-z) \phi(z') + g(z') \, \Bigr ] \, \Biggr \}_{z'=z} \non
\EA \quad\mbox{ trapezoidal rule} + \frac{h^2}{6} \phi_k - \frac{h^4}{120} \phi''_k  + {\cal O} 
\left ( h^6\right ) \> . 
\eea
where $B_{2m}$ are the Bernoulli numbers. Note that the correction terms come from discontinuities of 
derivatives of the integrands below and above the point 
$ z  = k \, h $ (we neglect the contributions around the endpoints at $\pm z_{\rm max}$ since the integrand 
is assumed to be very small there). The integration rule up to $ {\cal O} (h^2) $ is
denoted by ``Euler-MacLaurin$_2$'' in table \ref{tab: int_abs}.
Keeping correction terms up to $ {\cal O} (h^4) $ (``Euler-MacLaurin$_4$'') 
requires the second derivative of $\phi(z = k h)$ which we simply approximate by
\be
\phi''_k \E \frac{1}{h^2} \, \left [ \, \phi_{k-1} - 2 \phi_k + \phi_{k+1} \, \right ] + {\cal O} (h^2) \> .
\ee
Table \ref{tab: int_abs} shows that this gives vastly improved numerical results for the test functions: 
two or three orders of magnitude more accurate than the simple trapezoidal rule.
If there is another point at $ z' = 0 $ (as in the Green function of the ``ray'' representation or 
in the integrand of $I_2(z)$)
we simply add the corresponding correction for that point. 
% new 14. 12. 
For the variational calculation we used the  ``Euler-MacLaurin$_2$'' integration rule with 
$N = 60 - 120$ integration points.

\subsection{Variational equations}

We have solved the variational equations by iteration starting with the free solution. 
Updating during iteration is performed by the simple ``linear mixing scheme''
\be
y^{(n+1)}_{\rm in} \E \lambda_{\rm mix} \, y^{(n)}_{\rm out} + 
\left ( 1 -  \lambda_{\rm mix} \right ) \, y^{(n)}_{\rm in}
\ee
although more elaborate schemes are available (see e.g. ref. \cite{Baran} and references therein).
Simply taking $\lambda_{\rm mix} = 1 $ 
works quite well
for high-energy, forward scattering since the potential contribution is suppressed by appropriate 
powers of $K = k \cos \Theta/2$ in the ``aikonal'' representation, or $k$ in the ``ray'' 
representation. In the former case convergence obviously deteriorates at backward angles and one 
needs more and more iterations to fulfill the requirement that
\be
\left | \, \left ( \, X_0 + X_1 \, \right )^{(n+1)} -  \left ( \, X_0 + X_1 \, \right )^{(n)} \, 
\right | \> < \> \epsilon
\label{criterion}
\ee
which is imposed at fixed impact parameter $\fb$ before the iteration is allowed to stop. 
Typically, we take  
$ \epsilon = 10^{-5}$ or $ \epsilon = 10^{-6}$ which results in something like 
a dozen iterations for small $b$ whereas
only one iteration is needed for large (peripheral) values of the impact parameter 
because the potential
for those trajectories is already very feeble. The criterion (\ref{criterion}) makes sense because
the integrand is proportional to $ \exp (i (X_0 + X_1 + \ldots)) - 1$ so that considerable
computing time is saved for trajectories which barely feel the influence of the potential.
%\vspace*{0.4cm}

\subsection{Gaussian integrals and transforms}   %% 2. 7. 09
                                                 %% extended 27. 1. 10
\label{app: Gauss int}
Although in all textbooks the Gaussian integral in $n$ dimensions is given as
\bea
I_n(y) \EA \int d^n x \> \exp \left [ \, - x^T C x + i y^T x \, \right ] \non
\EA \frac{\pi^{n/2}}{\sqrt{\det_n C}} \,
\exp \left [ \, - \, \frac{1}{4} \, y^T C^{-1} y \, \right ] \> ,
\label{Gauss naive}
\eea
this strictly holds only if the matrix $C$ is real symmetric (or hermitean)
positive definite. Recall that this property implies positive eigenvalues  $\lambda_j$ so that
the determinant expressed as the product of the eigenvalues
\be
{\det}_n C \E \prod_{j=1}^n \lambda_j \> > 0
\ee
is positive and eq. \eqref{Gauss naive} is unambigous.
However, in our case
\be
C \E A + i B
\ee
is only {\it complex symmetric} (the quadratic form $ \> x^T C x \> $ projects out the symmetric part) and
for convergence of the multidimensional Gaussian integral
the real part $A$ has to be positive (semi-)definite, i.e. $ x^T A x \ge 0 $ for all $x$.
Under these conditions eq. \eqref{Gauss naive} still holds
provided the correct sign of the complex square root is chosen. This is a subtle but
important point about which we haven't found very much in the literature - except vague remarks that
``the sign of the square root is fixed by ... analytic continuation'' (ref. \cite{Wein}, p. 421).
A possible sign change
is equivalent to an additional phase $ \pi$ in the exponent of eq. \eqref{Gauss naive} and thus very similar
 to the Maslov phase correction (multiples of  $\pi/2$) for a semi-classical propagator 
when the trajectory of 
the particle goes through a focal point \cite{Maslov corr}.

In the following we outline how a proper treatment of these ``branch corrections'' may be obtained.
First, we specify which complex square root (which has branchpoints at $0$ and $\infty$)
we will use in the following: we choose the cut between these branchpoints 
along the negative real axis and define
the principal square root of a complex number $ z $ as the one with a positive real part, i.e.
\newcommand{\bsqrt}[1]{\sqrt[*]{#1}}
\be
\bsqrt{z} \Def \sqrt{ | z |} \, \exp \left ( \, \frac{1}{2} \, i \, {\rm arg} \, z \,
\right )
\> \> , \> \> |{\rm arg} \, z | < \pi \> .
\ee
This is also the value returned by the subroutine for the complex square root in the numerical implementation.
Then we assume that the complex symmetric matrix $C$ may be diagonalized by a complex orthonormal transformation
\begin{subequations}
\be
C \E O D O^T  \quad \quad {\rm with}\quad  O^T O \E O \, O^T \E 1 
\ee
and 
\be
D \E {\rm diag} \left (\lambda_1, \lambda_2 \ldots \lambda_n \right )
\ee
\end{subequations}
so that
\bea
I_n(y) \EA \int d^n x' \, \exp \left [ \, - \sum_{j=1}^n \lambda_j {x'_j}^2 +
i \sum_{j=1}^n \left ( y^T O \right )_j x'_j \, \right ] \non  
\EA \prod_{j=1}^n \left (  \int dx''_j \>
\exp \left [ \, - \lambda_j {x''_j}^2  - \frac{\left ( y^T O \right)_j^2}{4 \lambda_j} \, \right ] 
\right )
\label{Gauss ortho}
\eea
because the Jacobian of an orthonormal transformation is unity \footnote{Lacking an
explicit proof we here assume -- as done tacitly in all textbooks -- that
the complex orthonormal transformation of the original co-ordinates $x_i$ and the linear shift leads
to integration paths for the transformed co-ordinates $x_j''$ in the complex plane which can be safely
rotated back to the real axis. Analytic continuation of the real result faces a similar difficulty as
one can only perform it from a {\it region} in the complex plane and not from the real axis.}.
Due to a theorem by Bendixson (see eq. (6.9.15) in ref. \cite{StBu}) the
real parts of the complex eigenvalues
\be
{\rm Re} \> \lambda_j \> \ge \> 0
\ee
remain nonnegative when the eigenvalues of the real part of $C$ are assumed to be (semi-)positive definite.
In other words: we can write
\be
\lambda_j \E r_j \, e^{i \phi_j} \> \> , \> \> {\rm with} \> \> |\phi_j| < \frac{\pi}{2} \> .
\label{complex eigen}
\ee
Thus, each integral in the product of eq. \eqref{Gauss ortho} is convergent and we obtain
\bea
I_n(y) \EA \left ( \prod_{j=1}^n \> \bsqrt{\frac{\pi}{\lambda_j}} \right ) \, 
\exp \left [ \, - \frac{1}{4 \lambda_j} \,
\left ( y^T O \right)_j^2 \, \right ] \non
\EA \pi^{n/2} \, \left ( \prod_{j=1}^n \frac{1}{\bsqrt{\lambda_j}} 
\right ) \,
\exp \left [ \, - \frac{1}{4} \sum_{j=1}^n y_j ( O D^{-1} O^T )_j \,  y_j \, \right ] \non
\EA \pi^{n/2} \, \left ( \prod_{j=1}^n \,  \frac{1}{\bsqrt{\lambda_j}} \right ) \,
\exp \left ( \, - \frac{1}{4} y^T C^{-1} y \, \right )
\eea
as expected. However, it is important to note that the prefactor
is not given by the inverse square root of the
determinant of $C$ but by the product of the inverse square roots of the complex eigenvalues
\eqref{complex eigen}.
This may be different \footnote{Take the simple example: $ \> n = 3, \, \lambda_j \equiv \lambda = i + 0^+ $.
Then $\> \det_3 = - i - 0^+, \, \bsqrt{\det_3} = (1-i)/\sqrt{2}\> $ but $ \> (\bsqrt{\lambda})^3
= - (1-i)/\sqrt{2} \> $.}
depending on the value of
\be
\Phi^{(n)} \E \sum_{j=1}^n \phi_j \> \> , \> \> \> |\Phi^{(n)}| < n \frac{\pi}{2} \> .
\label{def phi_n}
\ee
Then
\bea
&& \prod_{j=1}^n \, \frac{1}{\bsqrt{\lambda_j}}  \E \frac{1}{\bsqrt{\det_n C}} \, \exp \left ( \, -
i \, n_{\rm br}(C) \, \pi \, \right ) \non 
&&\hspace*{-0.8cm} \E  \frac{1}{\sqrt{|{\det}_n C}|} \exp \left ( \, - \frac{i}{2}\,  
{\rm arg} \, {\det}_n C - i \, n_{\rm br} (C)\, \pi \, \right ) 
\eea
where the ``branch number'' is given by
\be
n_{\rm br} (C)  \E \left [ \, \frac{ |\Phi^{(n)}| + \pi }{2 \pi} \, \right ]
\ee
and $ [x] $ is the maximum integer not greater than $x$. \\$n_{\rm br} = 0, 2 \ldots $ denotes the principal
branch of the square root and $n_{\rm br} = 1, 3 \ldots $ the other, negative branch. From
eq. \eqref{def phi_n} we have the following bound
for a $(n \times n)$ complex symmetric matrix $C$ with nonnegative real part
\be
n_{\rm br} (C) < \left [ \, \frac{n + 2}{4} \, \right ] \> .
\label{n_br est}
\ee
Note that in contrast to the Maslov correction this
additional phase is not discontinous but rather corrects the phase jump when crossing the
(arbitrary) branch cut of the complex square root. In this way the analytic continuation of the result
\eqref{Gauss naive} is achieved. As an example take $\Phi = \pi + \epsilon$ to obtain
$-\frac{1}{2}{\rm arg} \> \det C  - n_{\rm br} \, \pi = -\pi/2 + |\epsilon|/2 $ for $\epsilon < 0
$ ($n_{\rm br} = 0$, above the cut) and
$ \pi/2 - \epsilon/2 - \pi = -\pi/2 - \epsilon/2 $ for $\epsilon > 0 $ ($n_{\rm br} = 1$, below the cut).

\vspace{0.2cm}

After these preliminaries we can now calculate the general Gaussian transform required in our variational
approach. This is particularly straight-forward and simple for a Gaussian potential: from
\bea
&& \tilde V_{\sigma}(\fp) \E \tilde V(\fp) \, \cdot \exp \left ( - \frac{1}{2} \, p_i \, \sigma_{ij} \,
 p_j \right )  \non
&& \hspace*{-0.8cm} \E V_0 \, \left (\pi R^2 \right )^{3/2} \, \exp  \left [ - \frac{R^2}{4} p_i \,
\left ( \delta_{ij} + 2 \sigma_{ij}/R^2 \right ) \, p_j \right ]
\eea
it follows by inverse Fourier transformation
\bea
V_{\sigma}(\fx) \EA  V_0 \, \pi^{3/2}  R^3 \, \int \frac{d^3 p}{(2 \pi)^3} \>
\exp  \left [ - \frac{R^2}{4} \, p_i \,  C_{ij} \, p_j  - i p_i x_i \right ] \non
\EA \frac{V_0}{\bsqrt{\det_3 C}} \, \exp \left [ \, - \alpha \, x_i \, C^{-1}_{ij} \, x_j
-i n_{\rm br}(C) \, \pi \, \right ]
\eea
where
\be
C_{ij} \E \delta_{ij} + 2 \alpha \, \sigma_{ij} \> , \> \> \> \alpha \E \frac{1}{R^2} \> .
\ee
is a complex symmetric $(3 \times 3) $-matrix whose real part should be positive (semi-)definite. 
Unfortunately
we were unable to verify this property analytically for our variational solutions but did not encounter
any numerical instabilities (which would be caused by a blow-up of $V_{\sigma}$)
during the iterative solution of the variational equations if $z_{\rm max}$
was not too large. Determinant and inverse of the $(3 \times 3)$-matrix $ C $ are known from elementary 
calculus
but we have not found a simple but reliable method to determine the ``branch number'' $n_{\rm br}$ which
in principle -- according to  eq. \eqref{n_br est} -- could be nonzero even in this case.
A method of ``branch tracking'' has been described in ref. \cite{ZoRe} but we used  a less elegant,
brute-force approach in which the complex eigenvalues were determined numerically with the NAG routine 
F02GBF and the prefactor was calculated as a product of the square roots of these.
As a by-product it was confirmed that the real parts of the eigenvalues
were always positive.
After evaluation of the Gaussian transform of the potential the Jacobian
and the Hessian then simply follow by differentiation:
\bea
\left ( J_{\sigma} \right ) _i & \equiv & \partial_i \, V_{\sigma}(\fx) \E - 2 \alpha
\left ( C^{-1} \right )_{ik} \, x_k \> V_{\sigma}(\fx) \\
\left ( H_{\sigma} \right )_{ij}  & \equiv & \partial_i \partial_j \, V_{\sigma}(\fx)
\E - 2  \alpha \, \Bigl [ \,  \left (   C^{-1} \right )_{ij} \\
&& \! \!  - 2 \alpha \,  \left (  C^{-1}  \right )_{ik} x_k \,  \left (  C^{-1}
\right )_{jl} x_l \, \Bigr ] \> V_{\sigma}(\fx) \> .
\eea
\vspace{0.1cm}

For the calculation of the second cumulant we also need the {\it double} Gaussian transform
\be
I_6  \Def \int d^3 p_1 \, d^3 p_2 \> \exp \left [ \, - \fp^T C \, \fp +  i \fx \cdot \fp \, \right ]
\label{double Gauss}
\ee
where
\be
C \E \left ( \begin{array}{cc}
              C_{11} & C_{12} \\
              C_{12} & C_{22}
              \end{array}\right )
\ee
is a  complex symmetric $(6 \times 6)$-matrix (see eq. \eqref{<chi2>}) and
\be
\fp \E \left ( \begin{array}{c}
              \fp_1 \\
              \fp_2 \end{array} \right )  \> , \>  \> \fx \E \left ( \begin{array}{c}
                                                      \fx_1 \\
                                                       \fx_2 \end{array}\right )
\ee
are 6-dimensional (column) vectors.
Determinant, inverse and branching number have been evaluated
as in the 3-dimensional case by calculating the complex eigenvalues -- a procedure which increased
the execution time of the program considerably. We found that (under our kinematical conditions)
no branch-crossing occured in the 3-dimensional calculation of the Gaussian transform of the potential
but the sign-change of the complex square root was essential to obtain the correct results in the
6-dimen\-sion\-al case (calculation of the second cumulant).
%\vspace*{1cm}

\subsection{Calculation of $\Omega$} 
Arising from  a functional determinant 
the quantity $\Omega$ poses a particular problem
for numerical evaluation. Several approaches are possible:
\vspace{0.1cm}

First, one may employ the classic method of Gel'fand and Yaglom \cite{GeYa} (already contained in 
textbooks, e.g. in ref. \cite{Schul}, Chapt. 6) to calculate a functional determinant as solution of an 
initial value differential equation. Indeed, defining
\be
\Omega_0 \Def \frac{1}{2} {\rm Tr} \, \ln \left ( 1 - \Sigma_0 H_{\sigma} \right )
\ee
(so that $\Omega = \Omega_0 + {\rm Tr} (\Sigma H_{\sigma})/2 $ )
its exponential is given  as ratio of two functional determinants
\be
\exp \left ( 2 \, \Omega_0 \right ) \E \frac{ {\rm Det} \left ( - \partial_t^2 - H_{\sigma}/m \right ) }
                                    {  {\rm Det} \left ( - \partial_t^2 \right ) } 
\label{Det GY}
\ee
and one may 
apply the Gel'fand-Yaglom procedure to evaluate it. However, the boundary conditions for the 
eigenfunctions $f(t)$ of the differential operators in eq. (\ref{Det GY}) are not of Dirichlet type 
as in the standard method but (in the ``aikonal'' case, cf. eqs. (\ref{bc 33 a}) and 
(\ref{bc 33 b}) ) of the form
\bea
\hspace*{-1cm} \lim_{T \to \infty} \left \{ \, f(T) + f(-T) - T \left ( \dot f(t) - \dot f(-T) \right ) 
\, \right \} \EA 0 \non
\lim_{T \to \infty} \left \{ \,  \dot f(t) + \dot f(-T) \, \right \} \EA 0 \> .
\label{bc Det}
\eea
Although Kirsten and McKane \cite{KMcK} recently have generalized the classic procedure 
to more general boundary conditions like (\ref{bc Det}) we do not follow this approach since 
simpler alternatives are available. 

%\twocolumn
%\vspace*{1cm}

%%%%%%%%%%%%%%%%%% Table test Omega (1. 7. 09), new: 2. 9. 09, extended: 10. 3. 10
\begin{table*}[htb]
\vspace*{0.5cm}
\caption{Comparison of values for the fluctuation term (functional determinant) $\Omega$ 
calculated in different ways
(see text) at $ b/R = 1$ for the Gaussian potential (\ref{Gauss pot}). 
Parameters of the calculation: $z_{\rm max}/R = 5 , \> (n_g,n_e) = (32,2) , \> 
\epsilon = 10^{-5} , \> \lambda_{\rm mix} = 1 $. Results for two values of the energy 
and the potential strength are displayed 
to allow comparison with the ``aikonal'' phases to which the total sum $X_0 + X_1 + i \, \Omega$ 
should tend in the high-energy limit.
}
\label{tab: test Omega}
\bce
\begin{tabular}{lcc|cc} 
 \hline\noalign{\smallskip}
                                          &                      &        &           &         \\
                                           &  \multicolumn{2}{c|}{$K R = 4 , \> 2 m V_0 R^2 = - 4$} &   
\multicolumn{2}{c}{$ K R = 8 , \> 2 m V_0 R^2 = - 8$}  \\
                                           &  \multicolumn{2}{c|}{( \# of iterations = 5 )} & 
 \multicolumn{2}{c}{( \# of iterations = 4 )} \\

                                           &                     &        &                   & \\
                                           &   Re          &  Im    & Re       & Im     \\
                                           &               &        &          &       \\
 \noalign{\smallskip}\hline\noalign{\smallskip}

                                           &               &        &          &        \\

 $\Omega$ , eq. (\ref{Om SH power})        &   -4.1931 (-3)  &   -6.2753 (-4) & -9.6244 (-4) &
                                                                                -4.3797 (-5)  \\
 $\Omega$ , eq. (\ref{Om S_0H power})   &   -4.1930 (-3)  &   -6.2751 (-4) & -9.6258 (-4) & 
                                                                                -4.3810 (-5)  \\
 $\Omega$ , eq. (\ref{Om mixed power})  &   -4.1935 (-3)  &   -6.2748 (-4) & -9.6243 (-4) &
                                                                                 -4.3812 (-5) \\
 $\Omega$ , eq. (\ref{Om SH})           &   -4.1934 (-3)  &   -6.2747 (-4) & -9.6246 (-4) & 
                                                                                -4.3814 (-5)\\
 $\Omega$ , eq. (\ref{Om mixed})       &   -4.1932 (-3)  &   -6.2756 (-4) & -9.6247 (-4) &
                                                                                -4.3799 (-5) \\
% new 10. 3. 10
 $\Omega$ , eq.  \eqref{Om eigen}      &   -4.1932 (-3) &    -6.2756 (-4) & -9.6247 (-4) &  
                                                                                -4.3799 (-5) \\  
                                    &                 &                &              & \\ \hline
                                    &                 &                &              & \\
         $X_0 + X_1$                &       3.4056 (-1)  &    8.1621 (-3) & 3.3382 (-1) & 
                                                                                1.8965 (-3) \\
     $X_0 + X_1 + i \, \Omega$      &       3.4118 (-1)  &    3.9688 (-3) & 3.3386 (-1) &
                                                                                9.3403 (-4) \\ 
                                    &                &                &              & \\\hline
                                    &                &                &              &   \\
$ \chi_{AI}^{(0)} +  \chi_{AI}^{(1)} +  \chi_{AI}^{(2 )}$  &     3.4176 (-1)  &    0.          
                                    & 3.3394 (-1)  &  0.       \\
$ i \, \omega_{AI}^{(2)}            $                &       0.           &    3.3216 (-3) & 0.  & 
                                                                               8.3041 (-4)  \\

\noalign{\smallskip}\hline

\end{tabular}
\ece
\vspace*{1.5cm}
\end{table*}
%%%%%%%%%%%%%%%% end table test Omega
%\vspace*{1cm}

\noindent
These include, second, the calculation of $\Omega$ as 
power series in either $ \Sigma H_{\sigma} $ or $ \Sigma_0 H_{\sigma} $:
\bea
\Omega \EA \frac{1}{2} \, {\rm Tr} \left [ \, - \ln \left ( 1 + \Sigma H_{\sigma} \right ) +  
\Sigma H_{\sigma} \, 
\right ] 
\label{Om SH}\\
\EA  \frac{1}{2} \, {\rm Tr} \, \sum_{n=2}^{\infty} \, \frac{(-)^n}{n} \, {\rm Tr} \left ( 
\Sigma H_{\sigma} \right )^n 
\label{Om SH power}\\
\Omega \EA \frac{1}{2} \, {\rm Tr} \left [ \, \ln \left ( 1 - \Sigma_0  H_{\sigma} \right ) +  
\frac{1}{ 1 - \Sigma^{(0)}  H_{\sigma} } - 1 \, \right ] 
\label{Om S_0H} \\
\EA  \frac{1}{2} \, {\rm Tr} \, \sum_{n=2}^{\infty} \, 
\left ( 1 - \frac{1}{n} \right ) \, {\rm Tr} \left ( \Sigma_0 H_{\sigma} \right )^n \> .
\label{Om S_0H power}
\eea
If the variational equations are fulfilled this should give identical results as should the 
``mixed'' form
\bea
\Omega \EA \frac{1}{2} \, {\rm Tr} \left [ \, \ln \left ( 1 - \Sigma_0  H_{\sigma} \right ) +  
\Sigma H_{\sigma} \, \right ] 
\label{Om mixed} \\
\EA  \frac{1}{2} \, \left [  \, - \sum_{n=1}^{\infty} \, \frac{1}{n} \, 
{\rm Tr} \left ( \Sigma_0 H_{\sigma} \right )^n + {\rm Tr} \left ( \Sigma H_{\sigma} \right ) \, \right ]  
\label{Om mixed power}
\eea
in which $ 1/(1 - \Sigma_0 H_{\sigma}) - 1 = \Sigma H_{\sigma} $ has been used. Note that
in the ``aikonal'' case the sum
also begins at $n = 2$ since $\Sigma_0(t,t) = 0$. If the variational equations are solved by
iteration it is consistent to evaluate the various sums in 
eqs. (\ref{Om SH power}), (\ref{Om S_0H power}) and (\ref{Om mixed power})
up to $ n = $ \# of iterations since each term is suppressed by an additional power of $K$ or $k$ --
if the iteration converges so will the power series expansion for $\Omega$. One should keep in mind 
that this procedure only works as long as $ K R \gg 1 $ or $ k R \gg 1 $, i.e. at high energies
and not too large scattering angles.
\vspace{0.2cm}

Third, one may evaluate the functional determinant as a $(3 \times n_g \times n_e)$-
dimensional ordinary determinant on the grid as one has discretized the time (or $z$-co-ordinate)
for the solution of the variational equations and the various integrals anyway. The NAG program
F03ADF was used for this purpose.

% new 10. 3. 10
Finally, the complex eigenvalues $\lambda_j$ of the matrix $ 1 + \Sigma H_{\sigma} $ may be 
calculated (with the help of the NAG routine F02GBF) so that from eq. \eqref{Om SH}
\be
\Omega \E \frac{1}{2} \, \sum_j \left [ \, \lambda_j - 1 - \ln \lambda_j \, \right ] \> .
\label{Om eigen}
\ee 
This also checks whether a ``branch crossing'' may have occurred which is
unlikely under these kinematic conditions as $\Sigma H_{\sigma}$ remains ``small'' and the power
series of the logarithm is well converging. Indeed, we found no case where the square root of 
the determinant was different from the product of the square roots of the eigenvalues. 
% end new 10. 3. 10  

\noindent
Table \ref{tab: test Omega} compares the results of the different methods in the ``aikonal'' 
representation at a fixed
value of the impact parameter. 
% changed 'good agreement' --> 'excellent agreement' 10. 3. 10:
One observes excellent agreement between the different methods.

Also included is a test at high energies where according to eqs. (\ref{X0+X1 HE}) and (\ref{Omega HE})
the phases and the fluctuation term can be described by the (much simpler) ``aikonal'' phases. 
%new 10. 12.
These have been worked out in  ref.  \cite{Carr} for spherically symmetric potentials 
(see eqs. (4.98), (4.99), (4.104) and (4.111) therein)
so that the corresponding expressions

%%%%%%%%%%%%%%%%%% Table FH test (11. 12. 09)
\begin{table*}[htb]
\vspace*{0.8cm}
\caption{Test of the Feynman-Hellmann relation (\ref{FH test}) for fixed $ \> b/R = 1 \> $ in 
the ``aikonal'' 
representation and fixed $ \> b_x/R = 0.6, b_y/R = 0.8 \> $ in the ``ray'' representation. In the 
latter case the scattering angle has been fixed at $\theta = 60^o$ with the momentum  transfer
along the $x$-axis. Accuracy parameters for the numerical solution of the variational equations are as in 
table \ref{tab: test Omega} and the integration over the potential
strength was performed by Gauss-Legendre integration with $n_{FH}$ points.
}
\label{tab: FH test}
\bce
\begin{tabular}{lcc|cc} 
\hline\noalign{\smallskip}
%                                           &                      &        &           &         \\

                                           &  \multicolumn{2}{c|}{``aikonal''} &   
\multicolumn{2}{c}{``ray''}  \\

                                           &   Re          &  Im    & Re       & Im     \\

\noalign{\smallskip}\hline\noalign{\smallskip}

 r.h.s. of eq. (\ref{FH test}) \quad $n_{FH} = \> \> 8 $  &   3.41182 (-1)  &   3.96843 (-3) & 2.50889 (-1) &
                                                                                3.38986 (-2)  \\
 r.h.s. of eq. (\ref{FH test}) \quad $n_{FH} = 12 $ &   3.41182 (-1)   &   3.96839 (-3) & 2.50889 (-1) &
                                                                                3.38987 (-2)\\
                                           &                 &                &              & \\ 
                                           &                 &                &              & \\

     $X_0 + X_1 + i \, \Omega$             &       3.41183 (-1)  &    3.96882 (-3) & 2.50889 (-1) &
                                                                                3.38993 (-2) \\ 

\noalign{\smallskip}\hline

\end{tabular}
\ece
\vspace*{2cm}

\end{table*}
%%%%%%%%%%%%%%%% end table FH test
for a Gaussian potential read
% end 10. 12.
read
\bea 
\chi_{AI}^{(0)} \EA - C \, \frac{\sqrt{\pi}}{2} \, e^{-y} \\
\chi_{AI}^{(1)} \EA - \frac{C^2}{K R} \,  \frac{1}{8} \sqrt{\frac{\pi}{2}} \, \left ( 1 - 4 y \right ) 
e^{-2 y} \\
\chi_{AI}^{(2)} \EA - \frac{C^3}{(K R)^2} \,  \frac{1}{16} \sqrt{\frac{\pi}{3}} \, \Biggl [ \, 1 - 
\left (12 + \sqrt{3} \pi \right ) y \non 
&& \hspace*{1.5cm} + \left ( 12 + 2 \pi/\sqrt{3} \right ) y^2 \, \Biggr ] \, e^{-3 y} \\
\omega_{AI}^{(2)} \EA - \frac{C^2}{(K R)^2} \,  \frac{\pi}{8} \, \left ( 1 - 4 y  + 2 y^2 \right ) \, 
e^{-2 y} \> , 
\eea
with $ C = 2 m V_0 R^2/(K R) $ and $ y = b^2/R^2 $.
\vspace*{0.2cm}

Good quantitative agreement is observed which becomes better at higher energies as expected. Note that
at $KR = 8$ the strength has also been changed to keep $ m V_0 /K  = $ constant. 
According to eqs. (\ref{X0+X1 HE}), (\ref{Im+Om HE}) the difference between
$X_0 + X_1 + i \, \Omega $ and $ \sum_{k=0}^2 \chi_{AI}^{(k))} + 
i \omega_{AI}^{(2)} $ then should decrease as $1/K^3$. 
Indeed, a closer look at the numerical values in table \ref{tab: test Omega} 
shows that this difference decreases by a factor $7.9$ in the real part and a factor $6.3$ 
in the imaginary part when doubling the energy which is in reasonable agreement 
with the expected factor $ (8/4)^3 = 8 $.

\newpage

\subsection{Test of the Feynman-Hellmann theorem}
\label{app: test of FH}

As explained in sect. \ref{sec: FH} the variational impact-parameter ${\cal S}$-matrix should fulfill 
additional relations since its ingredients are solutions of the variational equations. 
Here we will use the 
dependence (\ref{FH V0}) on the coupling strength $V_0$ to test our numerical solutions. Upon integrating
we should have
\be 
X_0 + X_1 + i \, \Omega \, \Bigr|_V \E \int_0^1 d\lambda \> \frac{1}{\lambda} \, X_0 \, 
\Bigr |_{V \to \lambda V} 
\label{FH test}
\ee
for fixed $\fb$ (and scattering angle). 
\vspace*{0.2cm}

The r.h.s. of this relation may be evaluated by simple
Gauss-Legendre integration (which has the advantage of avoiding the point $ \lambda = 0 $)
over the ''phase'' $X_0$ at different strength of the potential. As the integrand is a smooth function
of the potential strength very few Gaussian points are necessary to achieve a stable result 
which is in excellent agreement with the l.h.s. of eq. (\ref{FH test}). \\
This is displayed in 
table \ref{tab: FH test},
both for the ``aikonal'' and the ``ray'' representation and constitutes a rather stringent test 
that our numerical scheme for solving the variational equations and for calculating the stationary values 
is correct.

%%%%%%%%%%%%%
\vspace{1.5cm} 
{\it Note addded in proofs}. The correct mathematical framework to derive the Gaussian
integral with complex symmetric matrices is the theory of "pencils"   \cite{Boch} \quad
(called \\ "B\"uschel" in the German nomenclature \cite{Gant}) of quadratic forms which avoids the problems
indicated in footnote $^{10}$. Fortunately this approach leads to the same result and
procedures as used in Appendix D 4.

%%%%%%%%%%%%%%%%%%%%%%%%%%%%%%%%%%%%%%%%%%%%%%%%%%%%%%%%%%%%%%%%%%%%%%%%%
%%%% *************** REFERENCES *****************************************
%%%%%%%%%%%%%%%%%%%%%%%%%%%%%%%%%%%%%%%%%%%%%%%%%%%%%%%%%%%%%%%%%%%%%%%%

\end{document}